\documentclass[format=sigconf]{acmart}

\usepackage{algorithmic}
\usepackage{graphicx}
\usepackage{textcomp}
\usepackage{xcolor}
\usepackage{multirow}
\usepackage{booktabs}
\usepackage{siunitx}
\usepackage{stfloats}
\usepackage{caption}
\usepackage{subcaption}
\usepackage{braket}
\AtBeginDocument{%
  }

\newif\ifarxiv
\arxivtrue   

\providecommand{\ConfTitleSuffix}{}     
\providecommand{\ConfAuthorNote}{}     
\providecommand{\ConfHeaderNote}{}      
\providecommand{\ConfAcknowledgment}{}  

\ifarxiv
\else
  \IfFileExists{con_info.tex}{%
    \input{con_info.tex}%
  }{%
  }%
\fi

\ifarxiv
    \setcopyright{none}
  \renewcommand\footnotetextcopyrightpermission[1]{}
  \acmConference{}{}{}
  \acmBooktitle{}
  \acmDOI{}
  \acmISBN{}
\else
\ConfHeaderNote
\fi

\begin{document}

\newcommand{\liao} [1]{{\color{red} #1}}
\newcommand{\tokami} [1]{{\color{blue} #1}}

\ifarxiv
\title{Design automation and space-time reduction for surface-code logical operations using a SAT-based EDA kernel compatible with general encodings}

\else
\ConfTitleSuffix
\fi

\ifarxiv
\author{Wang LIAO}
\email{liao.wang@kochi-tech.ac.jp}
\orcid{0000-0003-2134-5588}
\affiliation{%
  \institution{Kochi University of Technology}
  \city{Kami}
  \state{Kochi}
  \country{Japan}
}

\author{Rei Tokami}
\email{tokami@qi.u-tokyo.ac.jp}
\orcid{0009-0009-2242-3673}
\affiliation{%
  \institution{The University of Tokyo}
  \city{Bunkyo}
  \state{Tokyo}
  \country{Japan}
}

\author{Yasunari Suzuki}
\email{yasunari.suzuki@riken.jp}
\orcid{0000-0002-8005-357X}
\affiliation{%
  \institution{Riken}
  \city{Wako}
  \state{Saitama}
  \country{Japan}
}
\fi

\ifarxiv
\renewcommand{\shortauthors}{ }
\else
\ConfAuthorNote
\fi
\begin{abstract}
Fault-tolerant quantum computers (FTQCs) based on surface codes and lattice surgery have been widely studied, and there is strong demand for a framework that can identify logical operations with low space-time cost, verify their functionality and fault tolerance, and demonstrate their optimality within a given search space, much like electronic design automation (EDA) in classical circuit design.
In this paper, we propose KOVAL-Q, an EDA kernel that verifies and optimizes surface-code logical operations by formulating them as a satisfiability (SAT) problem. Compared with existing SAT-based frameworks such as LaSsynth, our method can handle logical qubits with more flexible surface-code encodings, both as target configurations and as intermediate states. This extension enables the optimization of advanced layouts, such as fast blocks, and broadens the search space for logical operations. We demonstrate that KOVAL-Q can determine the minimum execution time of fundamental logical operations in given spatial layouts, such as $d$-cycle logical CNOTs and $2d$-cycle patch rotations. Their use reduces the execution time of widely studied FTQC applications by about 10\% under a simplified scheduling model. KOVAL-Q consists of three subkernels corresponding to different types of constraints, which facilitates its integration as a submodule into scalable heuristic frameworks. Thus, our proposal provides an essential framework for optimizing and validating core FTQC subroutines.
\end{abstract}



\maketitle
Quantum error correction~(QEC) is a vital technique for fault-tolerant quantum computing~(FTQC). Among the various approaches, encoding qubits with surface codes~\cite{kitaev1997,bravyi1998} and operating them via lattice surgery~\cite{horsman2012surface,fowler2018low} are particularly promising, as they can tolerate relatively high physical error rates and require only nearest-neighbor interactions. Optimizing such implementations involves placing surface-code qubits and routing logical operations through lattice surgery, which can be viewed as analogous to electronic design automation~(EDA) in classical circuit design.

A major challenge in this framework is that finding an implementation of fault-tolerant logical operations with minimum space-time cost is typically an intractable optimization problem. Therefore, establishing a framework for designing efficient and fault-tolerant logical operations on surface-code qubits is crucial for realizing practical FTQC. The exploration of efficient logical operations has been carried out either manually~\cite{fowler2012surface,litinski2019game,fowler2018low} or through simple algorithms tailored to a fixed device architecture, such as specific placements and encoding patterns of data logical qubits~\cite{beverland2022assessing,watkins2024high,kobori2025lsqca}. On the other hand, for frequently used subroutines such as a logical CNOT gate, it is desirable to perform a systematic search for near-optimal implementations, as is done in classical circuit design using EDA frameworks.

Recently, Tan \textit{et al.} proposed LaSsynth~\cite{tan2024}, a framework for finding efficient implementations of target logical operations by reducing the problem to satisfiability (SAT). In this method, the exploration space and constraints are defined using connected cubes in a three-dimensional (3D) representation of logical operations. Although this design choice can reduce the size of the resulting SAT instances, it also introduces two major limitations. First, this formulation cannot handle surface codes with general encoding schemes such as multi-qubit patches, multi-corner surface codes or fast block layout~(See Fig.~\ref{fig_sf_2q} for an example). This limits the applicability of this method because multi-qubit patches are used in several state-of-the-art FTQC implementations~\cite{litinski2019game,beverland2022assessing,silva2024multi,toshio2026star}. Second, the cube-based representation restricts the exploration space for optimization. Because it does not allow complicated qubit patches to appear as intermediate states, it may exclude optimal solutions. Therefore, as in classical circuit design, a versatile EDA framework is needed to unlock more efficient FTQC implementations.

To bridge this gap, we propose KOVAL-Q, an EDA Kernel for Optimizing and Verifying lattice surgery implementation with Arbitrary surface-code encoded Logical Qubits.
Compared with LaSsynth~\cite{tan2024}, the key difference is that KOVAL-Q expresses the exploration space and constraints in terms of stabilizer faces rather than cubes. This enables optimization and verification for general classes of surface-code encodings, including multi-qubit patches, albeit at the cost of larger SAT instances. KOVAL-Q consists of three subkernels: Lattice Surgery SAT (LS-SAT), which formulates legal lattice-surgery operations; Functional SAT (Func-SAT), which verifies the equivalence of quantum gate operations across candidate implementations; and Fault-Tolerant SAT (FT-SAT), which checks the fault tolerance of an implementation. These subkernels can be used either individually or in combination, making KOVAL-Q a reusable foundation for future FTQC EDA toolchains.

We demonstrate the advantages of our method from two perspectives. First, we show that it can automatically discover fundamental logical operations on two-qubit surface-code patches. To the best of our knowledge, no previous framework has been able to systematically discover and optimize such operations. Second, we show that our formulation is beneficial even for optimizing logical operations on single-qubit patches, i.e., standard surface-code patches. We find that a logical CNOT gate and patch rotation can be implemented with $d$ and $2d$ stabilizer measurement cycles, respectively (see Figs.~\ref{fig_1-beat_cnot} and \ref{fig_2-beat_rotation}). To the best of our knowledge, these two solutions improve the minimum latency of the logical operations and, importantly, lie outside the exploration space of the existing formalism. According to our simplified estimate, the use of them reduces the execution times of random Clifford circuits and FTQC applications by 22\% and about 10\%, respectively. These results clearly illustrate the significance of our proposal.

Although KOVAL-Q generates larger SAT instances than LaSsynth, this is an acceptable trade-off because it enables improvements in fundamental subroutines such as CNOT and can therefore serve as a complement to existing frameworks. We also emphasize that, when the goal is verification rather than optimization, KOVAL-Q can handle large instances as well. Although our validation in this paper focuses on surface codes defined on a grid lattice, the underlying SAT formulation is expected to be broadly applicable to other topological QEC codes, such as color codes~\cite{landahl2011fault,lacroix2025scaling}. Thus, our proposal provides an indispensable framework for optimizing and validating core FTQC subroutines.

To summarize, the contributions of this work are as follows:
\begin{itemize}
\item We provide an EDA kernel, KOVAL-Q, for optimizing and verifying logical operations compatible with general surface-code encodings. To this end, we formulate the constraints of lattice surgery on surface codes in terms of stabilizer faces.
\item Our EDA kernel consists of three separable subkernels for verifying different classes of constraints in lattice-surgery operations, thereby enabling flexible use in a wide range of FTQC EDA tasks.
\item We demonstrate the proposed EDA kernel by discovering logical operations on two-qubit patches and by performing optimization over an expanded exploration space. In particular, we present a logical CNOT gate and patch rotation that require $d$ and $2d$ repetitions of stabilizer measurements.
\end{itemize}

\section{Background}
\subsection{Surface code}
Physical qubits, the fundamental units of quantum computing, are highly sensitive to noise. For example, the lifetime of superconducting-circuit qubits is limited to about $1~\mathrm{ms}$~\cite{chow2021,google2023,google2025quantum}. In contrast, solving practical problems requires maintaining quantum states for hours~\cite{babbush2018,yoshioka2024hunting}. QEC bridges this gap by encoding logical qubits into blocks of physical qubits and detecting/correcting errors through a set of Pauli measurements.

Among many QEC codes, surface codes are well-studied QEC codes, as they offer high error-correction performance and can be implemented using only nearest-neighbor interactions on qubits placed on a two-dimensional grid. Most current experimental demonstrations and architecture designs assume surface codes as a standard choice~\cite{google2023,google2025quantum}.

\begin{figure}[t]
  \centering
  \begin{subfigure}{0.39\columnwidth}
    \includegraphics[width=1\columnwidth]{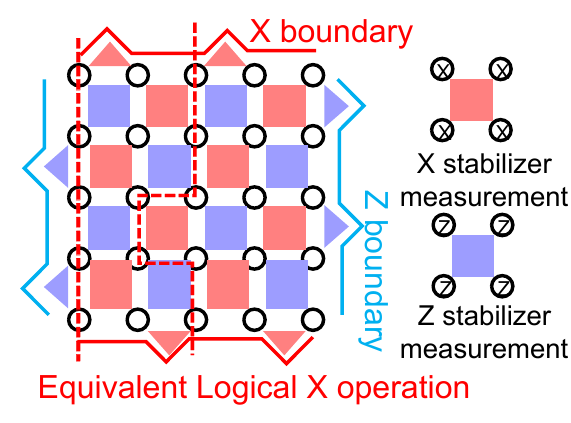}
  \vspace{-10pt}
    \caption{one logical qubit patch}
    \label{fig_sf_1q}
\end{subfigure}\hspace{20pt}%
  \begin{subfigure}{0.4\columnwidth}
    \includegraphics[width=1\columnwidth]{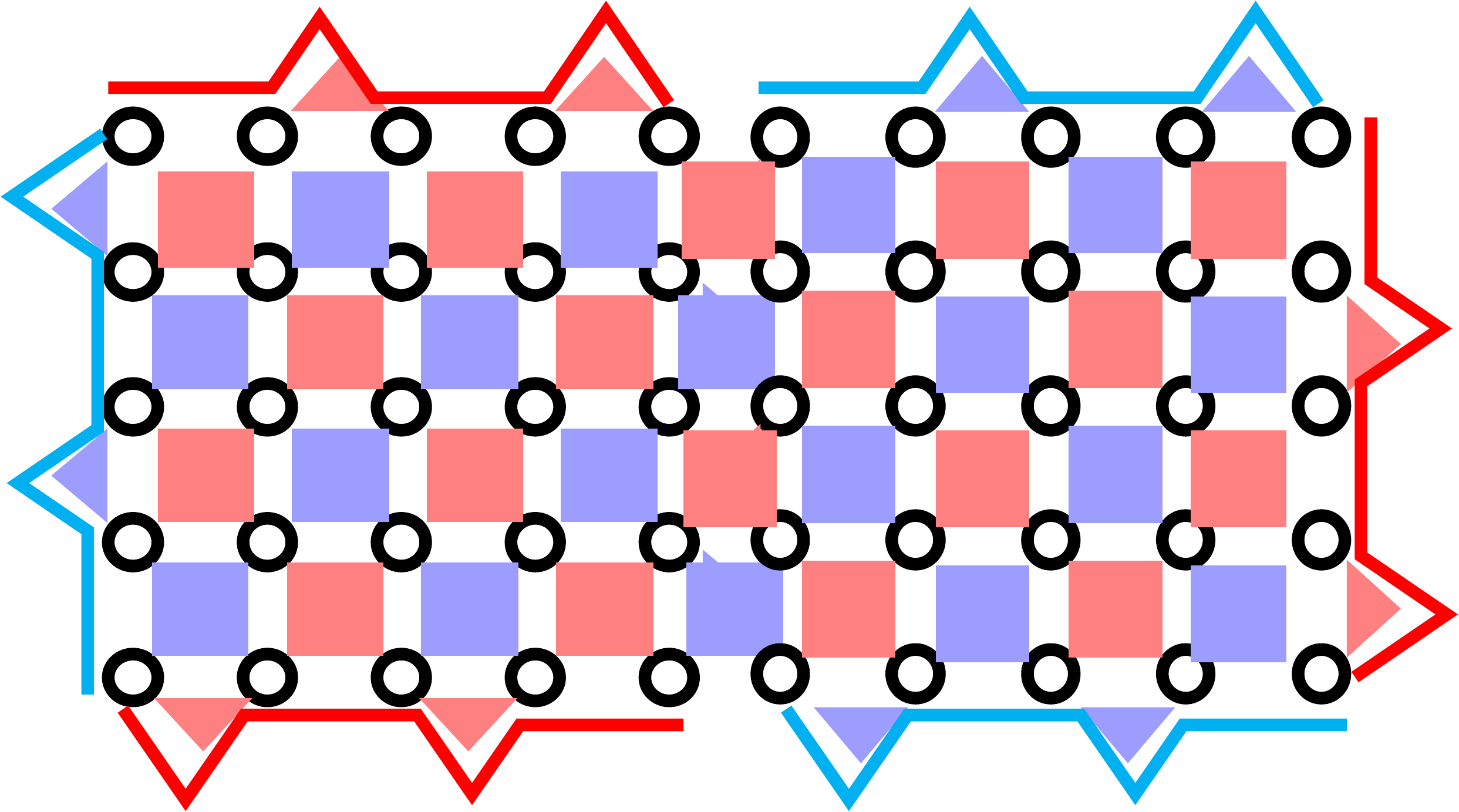}
  \vspace{-10pt}
    \caption{two logical qubit patch}
    \label{fig_sf_2q}
  \end{subfigure}
    \vspace{-10pt}
  \caption{Surface-code logical qubits with boundaries}
  \label{fig:sf_patch}
  \vspace{-15pt}
\Description{}
\end{figure}

Here, we briefly explain the properties of surface codes. For details, see Refs.\,\cite{fowler2012surface,fowler2018low,krinner2022}. 
A logical qubit encoded with surface codes is represented by a cell of physical qubits as shown in Fig.\,\ref{fig_sf_1q}. White circles represent physical qubits that form an encoded quantum state. The red and blue polygons represent parity-check patterns of QEC codes, as shown on the right. We measure the parity of two types of quantum errors, phase flip and bit flip, on physical qubits located at the corners, respectively. These measurements are called stabilizer measurements, and we can track errors from observed parity values by repeating them. The period of stabilizer measurements is referred to as a code cycle.
There are two types of boundaries in qubit patches: $X$ and $Z$ boundaries. A chain of Pauli errors connecting two distinct $X$ ($Z$) boundaries forms a Pauli $X$ ($Z$) logical operation on encoded qubits. The width of a surface-code patch, measured in code-cell units, corresponds to the code distance $d$ of the surface code.

In general form, surface codes can be used to encode multiple qubits with multiple patches. The number of distinct boundaries determines the number of encoded qubits within the patch. Specifically, a patch with $k$ $X$ and $Z$ boundaries encodes $k-1$ logical qubits. For example, Fig.\,\ref{fig_sf_2q} shows a two-qubit patch.

\subsection{Lattice surgery}
To execute an arbitrary quantum circuit fault-tolerantly, we need a universal set of logical operations acting on surface-code logical qubits. Code deformation~\cite{bombin2009quantum} provides a powerful framework for implementing universal Clifford logical operations only with nearest-neighbor interactions. In this framework, logical operations are executed fault-tolerantly by sequentially modifying the stabilizer measurement pattern, thereby gradually transforming the logical state. When part of the stabilizer measurement pattern is changed, the corresponding parity-check outcomes become random. Therefore, the sequence must be carefully designed so that any non-trivial physical error acting on fewer than $d$ physical qubits remains detectable, where $d$ is a code distance.

A representative example of code deformation is lattice surgery~\cite{horsman2012surface}, in which the parity-check patterns of two patches are merged, maintained for $d$ rounds of stabilizer measurements, and then split again. This procedure yields a fault-tolerant logical Pauli measurement in the $XX$ or $ZZ$ basis, depending on whether the connected boundaries are of $X$-type or $Z$-type. Throughout this paper, we use the latency of lattice surgery, i.e., $d$ rounds of stabilizer measurements, as a unit of logical operations, called a code beat.

\begin{figure}
    \centering
    \includegraphics[width=0.85\linewidth]{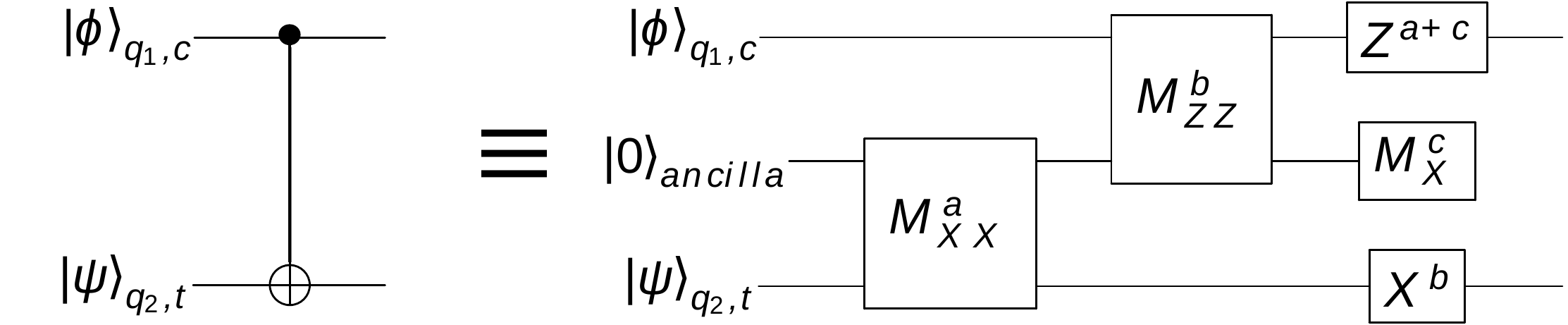}
    \vspace{-5pt}
    \caption{Lattice-surgery-based CNOT gate}
    \vspace{-20pt}
    \label{fig_cnot_ls}
\Description{}
\end{figure}

Using code-deformation techniques such as lattice surgery, various logical Clifford gates can be constructed. For example, a CNOT gate can be implemented using lattice-surgery sequences, as shown in Fig.\,\ref{fig_cnot_ls}~\cite{fowler2018low}. By combining these operations with noisy magic-state preparation, one can obtain a universal set of logical operations. 

Although Fig.\,\ref{fig_sf_1q} shows a variant of the surface code known as the rotated surface code~\cite{fowler2018low,krinner2022}, the same discussion applies to the standard surface code~\cite{bravyi1998}. Therefore, throughout the discussion below, we do not distinguish between these surface-code variants.

\section{The construction of KOVAL-Q}
\label{sect_kernel}

\subsection{Exploration domain}
\begin{figure}[t]
  \centering
  \begin{subfigure}{0.45\columnwidth}
  \includegraphics[width=\linewidth]{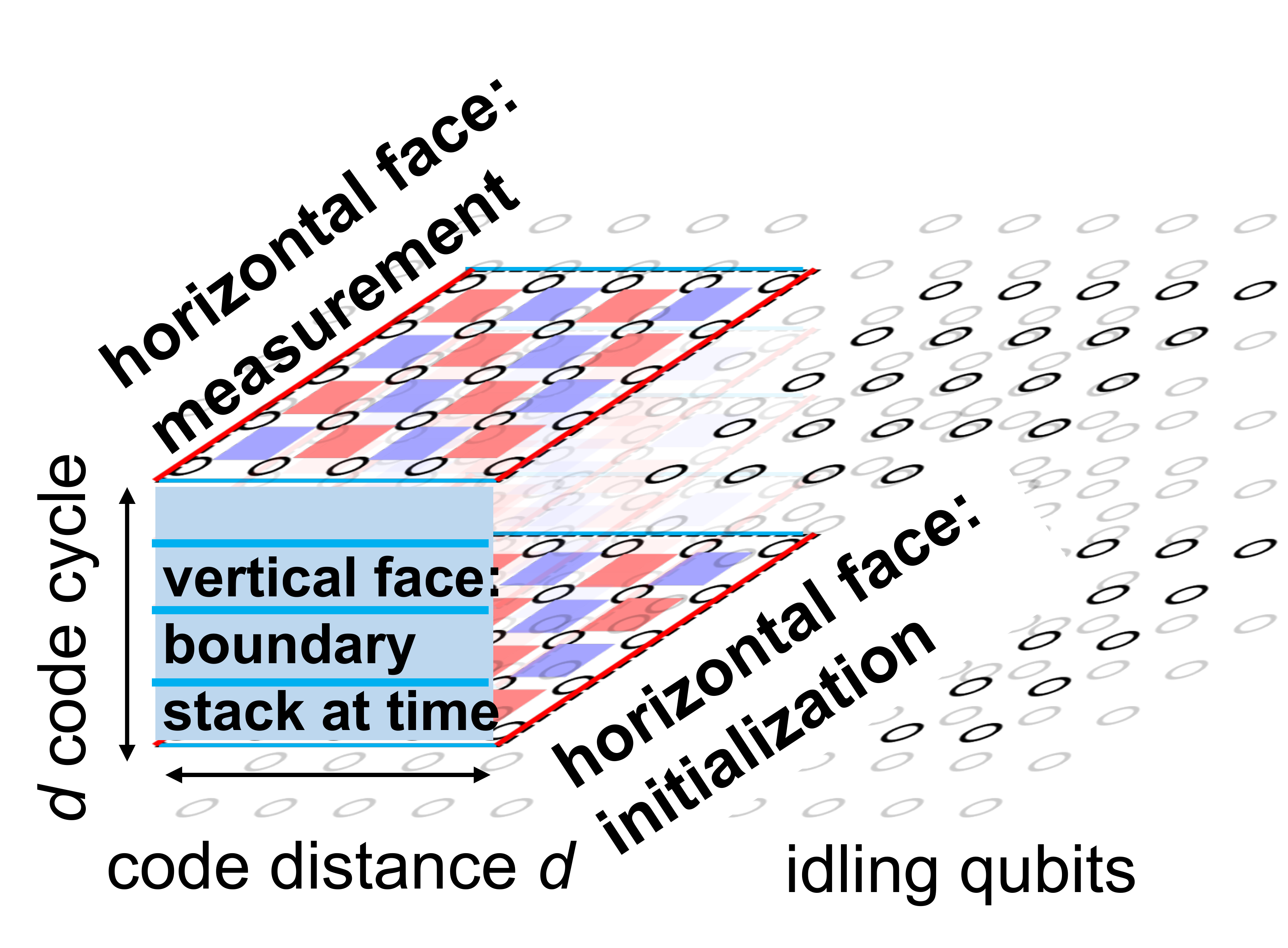}
    \caption{Stabilizer face definition}
    \label{fig_stab_faces}
  \end{subfigure}\hfill%
  \begin{subfigure}{0.55\columnwidth}
    \includegraphics[width=\linewidth]{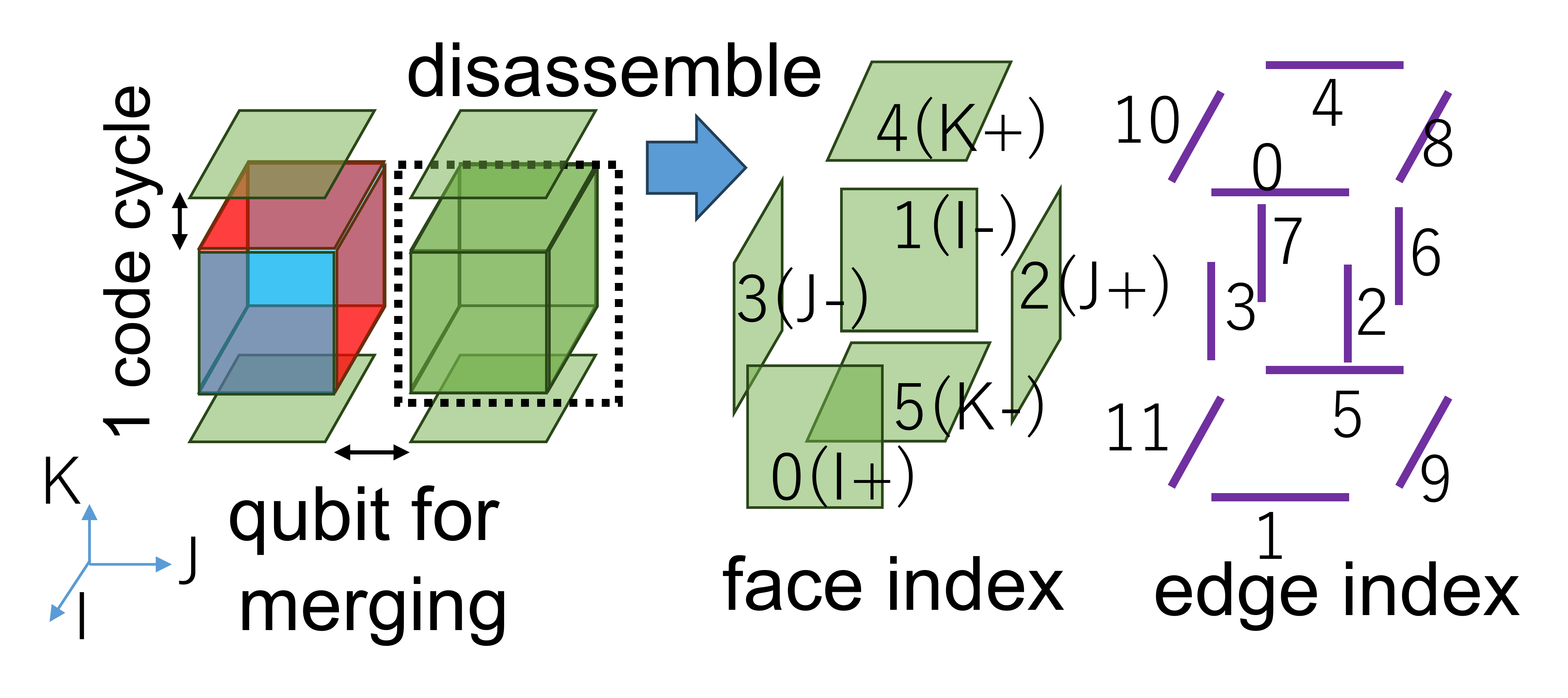}
    \caption{Exploration domain and indices}
    \label{fig_face_assembly}
  \end{subfigure}
  \vspace{-22pt}
  \caption{Definition of stabilizer faces and exploration domain}
  \label{fig:all-faces}
  \vspace{-15pt}
\Description{}
\end{figure}

To design a fault-tolerant logical operation, we must ensure:
(1) the overall action on the code space is identical to the target logical operation in the absence of error; and
(2) the procedure detects any nontrivial physical error chain of weight less than $d$.
For surface codes, we can judge these two conditions using the following three-dimensional~(3D) representation. We consider a 3D domain where the $IJ$-plane represents qubit resources on a device and the $K$-axis represents machine time. In each code cycle, the $X$ and $Z$ boundaries of qubit patches are drawn as blue and red edges, respectively. Repeating the same stabilizer measurements for $d$ cycles produces a 3D bulk with height $d$ bounded by red and blue faces.
Such a face, accumulated from boundaries, is defined as a stabilizer face, as shown in Fig.\,\ref{fig_stab_faces}. 
Figure\,\ref{fig_face_assembly} shows the index of faces and their boundaries (edges) for later reference, as we will use them as an exploration domain.
Note that, since $d$ becomes the unit of domain space, the algorithmic complexity of KOVAL-Q is independent of the code distance $d$.

KOVAL-Q verifies the fault tolerance and logical actions of given physical operations by forming and solving a SAT instance with three sets of constraints.
The first set of constraints is LS-SAT, which verifies that the connectivity of the boundaries is valid as lattice surgery.
With the framework of stabilizer flows~\cite{aaronson2004improved}, these relations must coincide with those derived from the action of logical operations, which are validated by the second set, Func-SAT. A sufficient condition for fault tolerance is rephrased as the absence of any chain that satisfies all of the following conditions: 1) the chain is inside the 3D object; 2) the length of the chain is less than $d$; and 3) the chain connects two unconnected faces with the same color. The last set, FT-SAT, verifies them.
In summary, finding an efficient logical operation is equivalent to finding a colored 3D bulk satisfying these conditions while minimizing its volume.

Next, we present the formulation of the KOVAL-Q kernel, which generates SAT instances for finding and verifying the FTQC implementations.
The details of each subkernel are described below.

\subsection{Variables and constraints of LS-SAT}
\label{subsec:LS-SAT}
\begin{figure*}
    \centering
    \begin{subfigure}{0.25\linewidth}
    \includegraphics[width=1\linewidth]{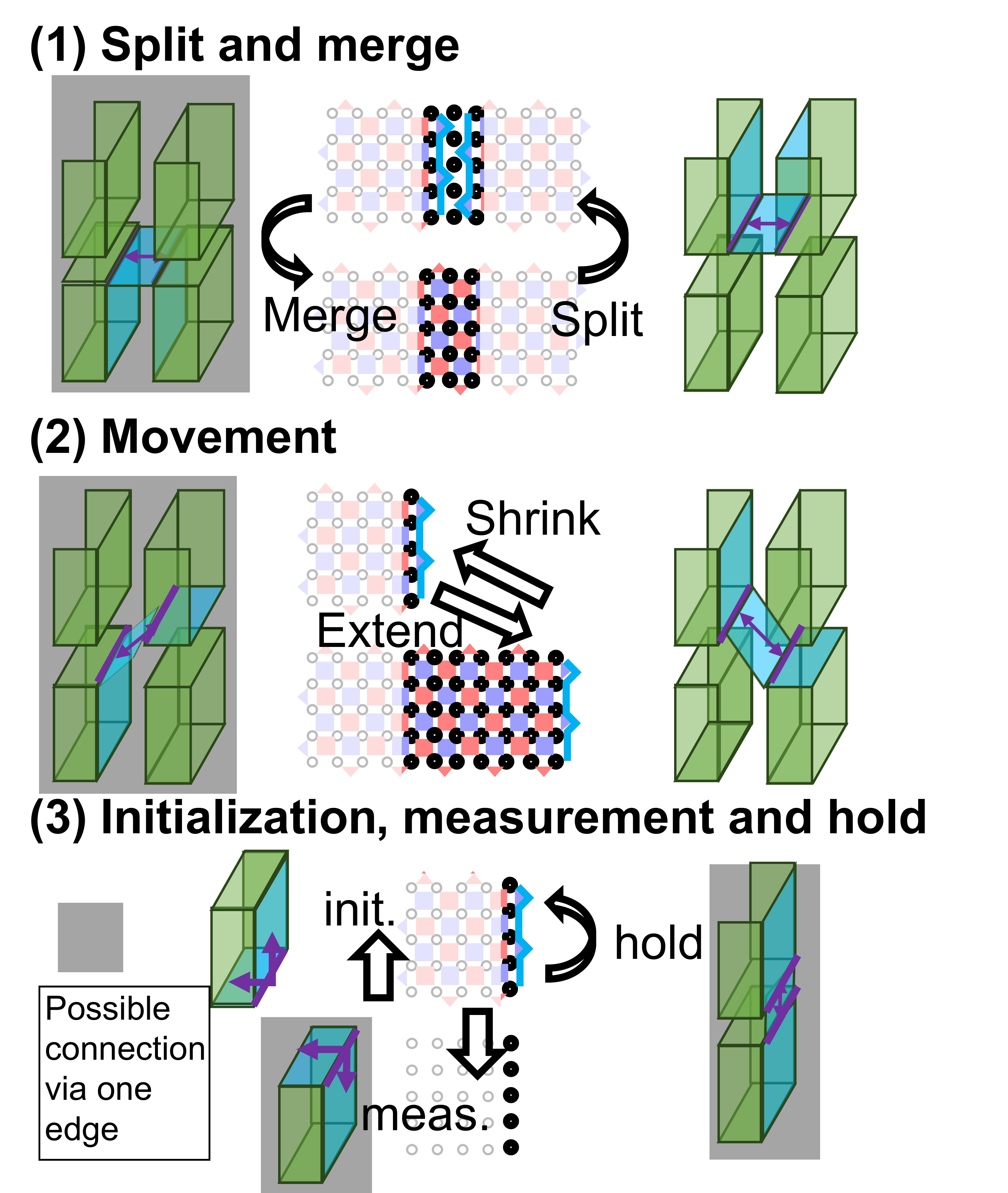}
    \caption{Lattice-surgery operations}
    \label{fig_ls_operations}
    \end{subfigure}%
    \begin{subfigure}{0.32\linewidth}
    \includegraphics[width=1\linewidth]{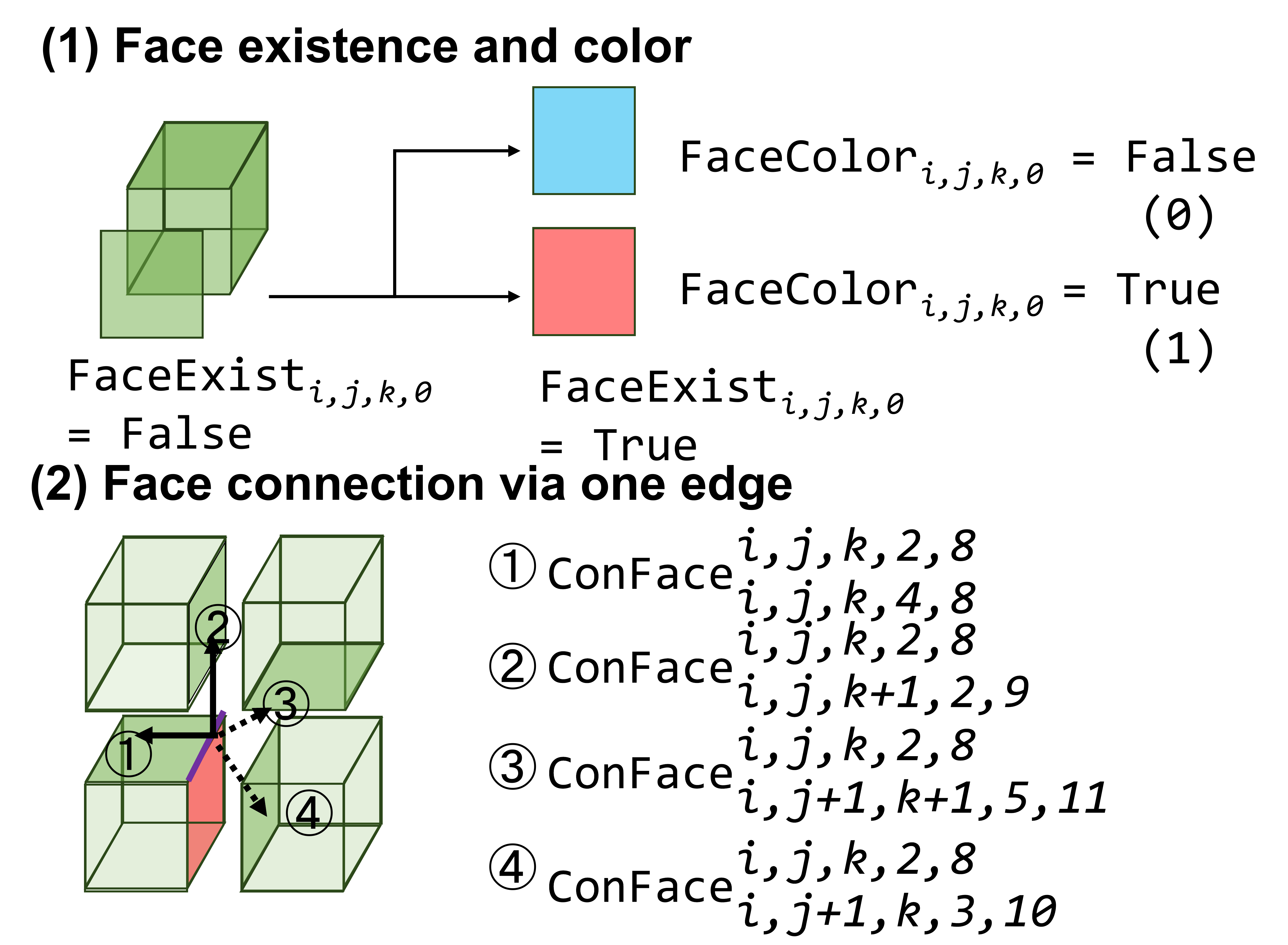}
    \caption{Variables of faces}
    \label{fig_lssat_vars}
    \end{subfigure}%
    \begin{subfigure}{0.35\linewidth}
    \centering
    \includegraphics[width=1\linewidth]{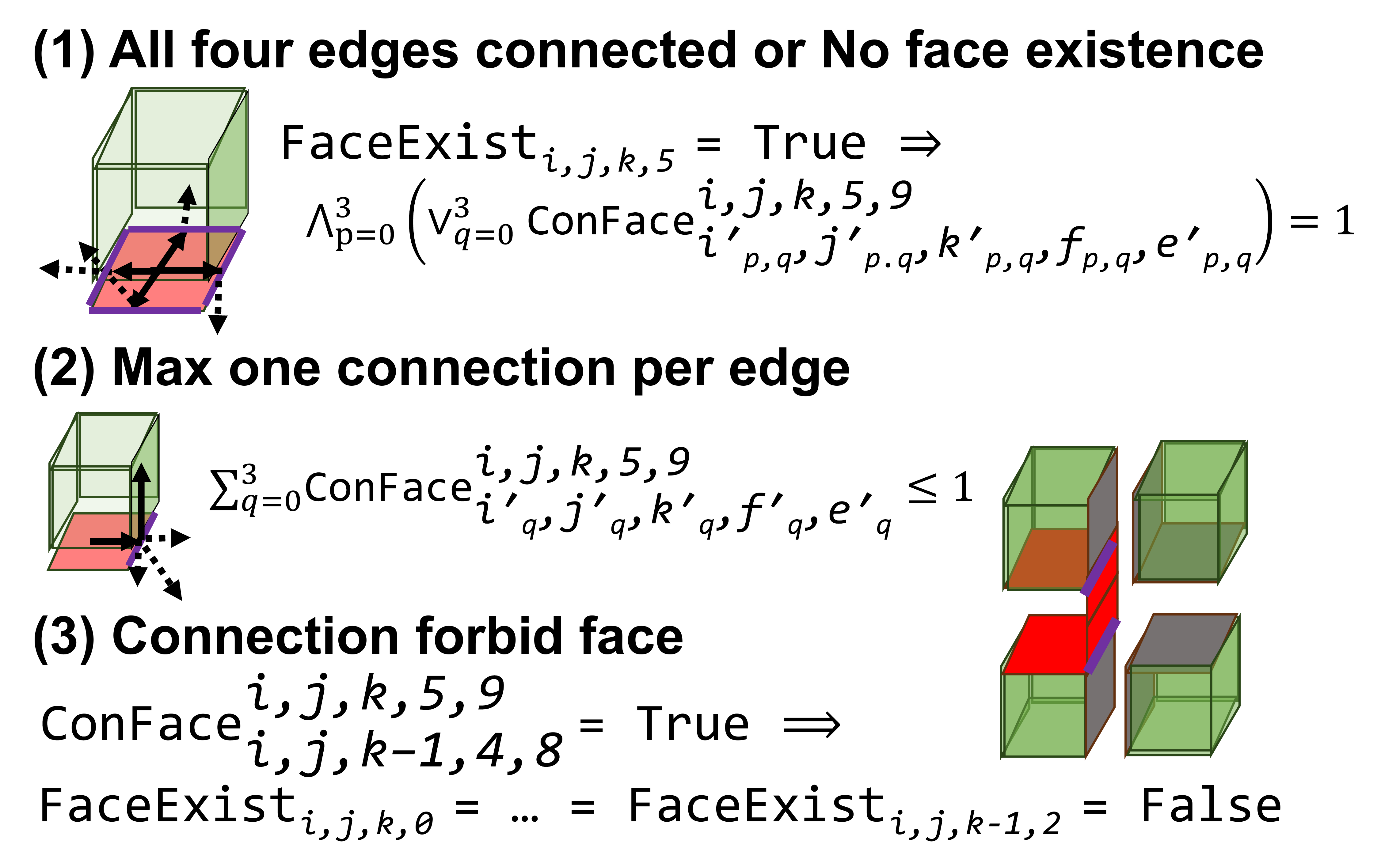}
    \caption{Constraints for LS-SAT}
    \label{fig_lssat_cons}

\end{subfigure}
\vspace{-10pt}
\caption{Topological expression of lattice-surgery operations, variables and constraints for LS-SAT}    \vspace{-10pt}
\Description{}
\end{figure*}

Within the exploration domain, we formulate LS-SAT constraints to check whether a 3D bulk of stabilizer faces constitutes a valid sequence of lattice-surgery operations.
The idea is straightforward: emulate all the possible lattice-surgery operations and re-express them in terms of topological connections between stabilizer faces.
When stabilizer measurement patterns change within a cycle due to lattice surgery, the cross-section in the $IJ$-plane also changes. 
The colors of the bottom (top) faces of the 3D bulks that appear when expanding (shrinking) stabilizer measurement patterns are determined by the initialization (measurement) basis of the added (removed) physical qubits. 
For detailed rules and theories, please see Refs.~\cite{fowler2018low,tan2024}. Figure\,\ref{fig_ls_operations} depicts the relations between stabilizer-face connections and lattice-surgery operations, consisting of the above-mentioned operations. This topological description applies uniformly to all boundaries located at the edges of horizontal faces.

To describe such connection in SAT formulas, we assign the Boolean variables to each stabilizer face at location $(i,j,k)$ with face index $f$: $\mathtt{FaceExist}_{i,j,k,f}$ for the existence of face, $\mathtt{FaceColor}_{i,j,k,f}$ for colored Pauli type ($X$ or $Z$), $\mathtt{ConFace}^{i,j,k,f,e}_{i',j',k',f',e'}$ for connection of face $f$ via edge $e$ to edge $e'$ of face $f'$ at location $(i',j',k')$.
Each face connects through one edge, indexed by $e$. A valid connection has four possible candidates as illustrated in Fig.\,\ref{fig_lssat_vars}(2). 
Connection to the other four neighbor faces is not allowed since they result in an illegal lattice-surgery operation, e.g., $\mathtt{ConFace}^{i,j,k,2,8}_{i,j,k+1,5,9}$ (K- shown in Fig.\,\ref{fig_face_assembly}, bottom face of cube at beat $k+1$) would represent boundary disappearance after initialization.
These restrictions stem from the topological consistency, i.e., each face implicitly has a normal vector that cannot be reversed.

To ensure that the faces and connections of the generated 3D bulk are valid, we impose the LS-SAT constraints in Fig.\,\ref{fig_lssat_cons}. 
The first constraint shown in Fig.\,\ref{fig_lssat_cons}(1) is that a face exists if and only if all four of its edges are connected to other faces.
This constraint is expressed as:
$$
\lnot \mathtt{FaceExist}_{i,j,k,f} \vee \bigwedge_{p=0}^3 \bigvee_{q=0}^3  \mathtt{ConFace}^{i,j,k,f,e_{p}}_{{i_{p,q}',j_{p,q}',k_{p,q}',f_{p,q}',e_{p,q}'}},
$$
where $p$ indexes edges on the face, and the $q$ enumerates candidate connections via $p$ edge.
This applies to all the faces in the exploration domain.

The second constraint is that each edge can have at most one connection, as shown in Fig.\,\ref{fig_lssat_cons}(2).
$$
\sum_{q = 0}^3 \mathtt{ConFace}^{i,j,k,f,e}_{i_q',j_q',k_q',f_q',e_q'} \le 1
$$
This ensures that new stabilizer boundaries are introduced only through valid operation.

The third constraint ensures that shrinking and re-expanding a patch does not reintroduce forbidden stabilizer boundaries (Fig.\,\ref{fig_lssat_cons}(3)). For example, when a patch shrinks in the $J$ direction and expands back, the stabilizer boundary must remain absent inside the patch. This is enforced as follows.
$$
(\lnot \mathtt{ConFace}^{i,j,k,f,e}_{i', j', k', f', e'} \lor (\hspace{-15pt}\bigwedge_{\delta\in\Delta(f, e, fi')}\hspace{-15pt} \lnot \mathtt{FaceExist}_{(i,j,k,f)+\delta})))
$$
Here, $\Delta$ specifies the offset positions of forbidden faces relative to the current face of $(i,j,k,f)$.
Together, these constraints ensure that SAT solutions correspond to a legal sequence of lattice-surgery operations consistent with code-deformation rules.

\subsection{Variables and Constraints of Func-SAT}
\label{subsect_funcsat}

\begin{figure*}
    \centering
    \begin{subfigure}{0.23\linewidth}
  \centering
  \raisebox{0.1\height}{\includegraphics[width=\linewidth]{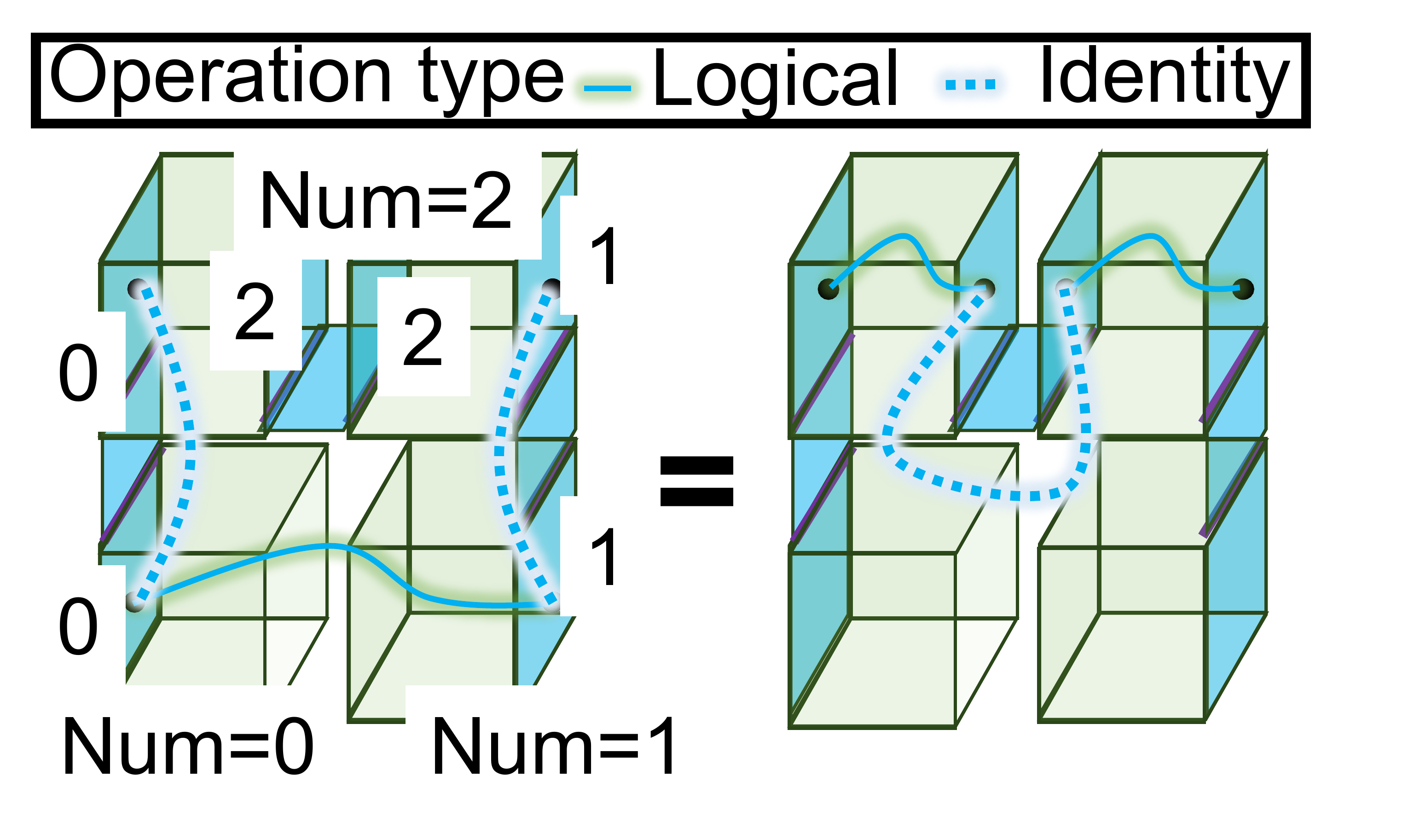}}
  \caption{Stabilizer flow equivalence reasoning by lattice-surgery operation}
  \label{fig_funcsat_split}
\end{subfigure}\hspace{10pt}%
    \begin{subfigure}{0.27\linewidth}
    \includegraphics[width=1\linewidth]{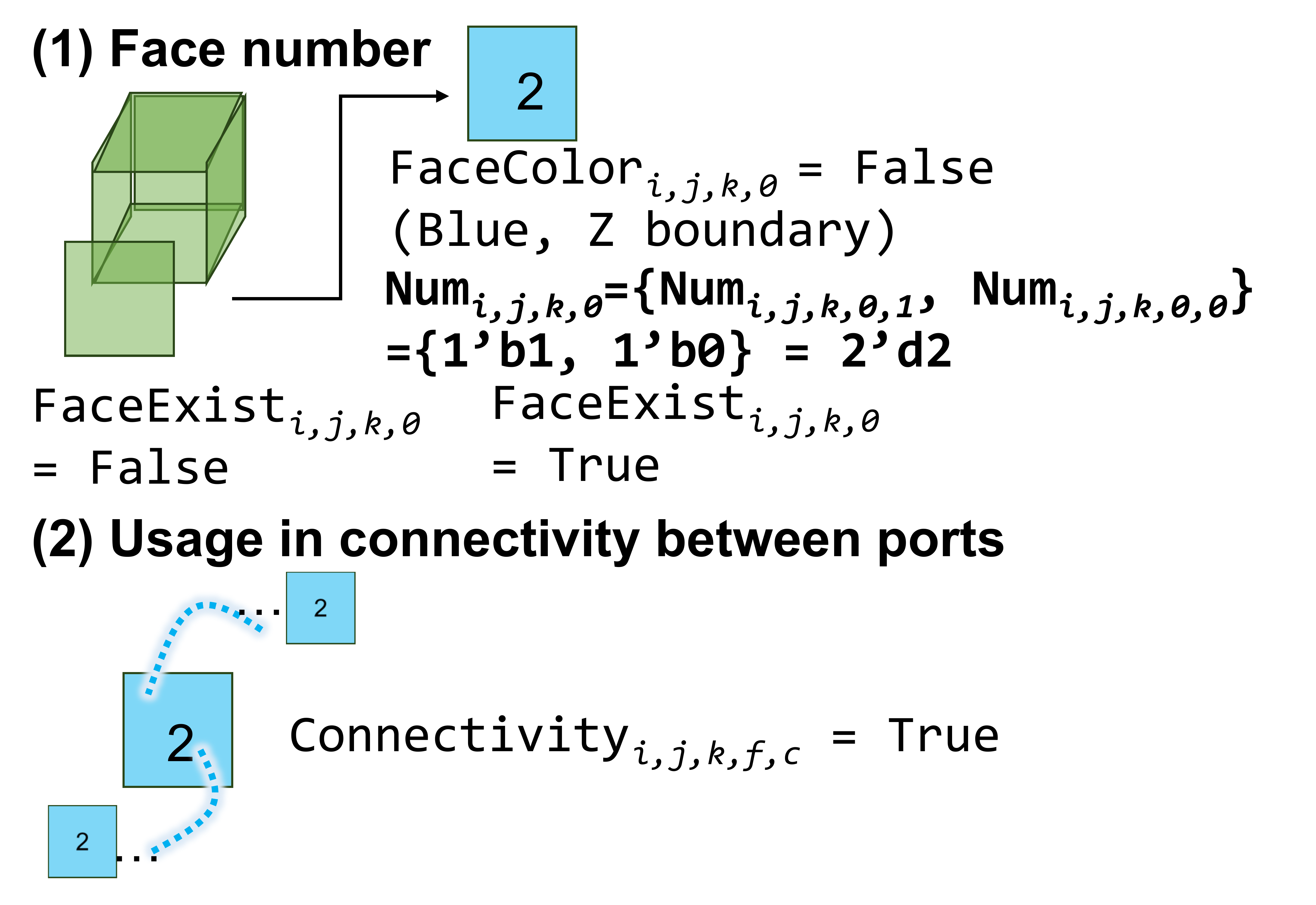}
    \caption{Variables of Func-SAT}
    \label{fig_funcsat_facenum}
    \end{subfigure}%
    \begin{subfigure}{0.35\linewidth}
    \includegraphics[width=1\linewidth]{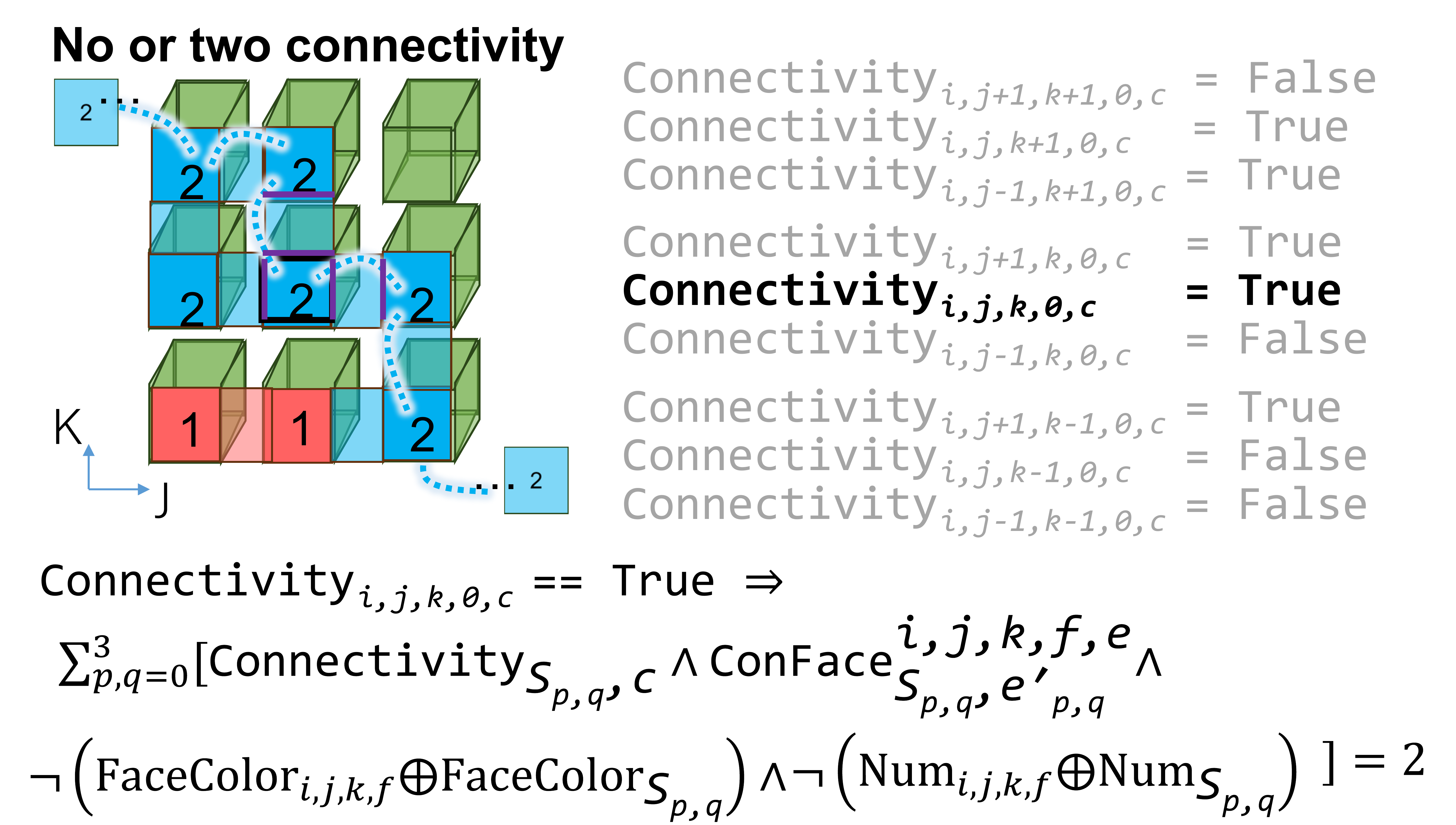}
    \caption{Constraints of Func-SAT}
    \label{fig_funcsat_connectivity}
\end{subfigure}
\vspace{-10pt}
\caption{Topological expression of stabilizer flow, variables, and constraints for Func-SAT formula}
\vspace{-10pt}
\Description{}
\end{figure*}

Next, we define variables and constraints for Func-SAT to validate the equivalence between the target action and that of a generated 3D bulk.
For this purpose, we employ the idea of stabilizer flow~\cite{tan2024}, which can be used for logical Clifford circuits. This framework enables the equivalence check by tracking the transformation of logical Pauli operators in the 3D domain representation.
Here, we briefly explain the idea of stabilizer flow. Given a target logical Clifford $C$, the action of a logical Pauli operator $P$ before $C$ is equivalent to $CPC^{\dagger}$ after $C$ (i.e., $CP = (CPC^{\dagger}) C$). Thanks to the property of the Clifford group, the transformed $CPC^{\dagger}$ remains as a logical Pauli operator and can be evaluated efficiently~\cite{aaronson2004improved}. Moreover, two processes are equivalent if this mapping from $P$ to $CPC^{\dagger}$ is identical. Thus, we can validate the equivalence of two processes by comparing the transformation of logical Pauli operations.

It is trivial to find the transformation of a given logical Clifford action $C$. Thus, what we need to find is the transformation of logical Pauli operators by a 3D object. This can be represented as constraints as follows. A logical $X$ (or $Z$) Pauli operator in surface codes is a chain of physical bit (or phase) flips connecting distinct faces with the same color in the 3D object. Since a closed chain of errors inside the 3D object and that connects the connected faces with the same color are identical operations on the code space, we can continuously transform the chain of logical Pauli operators along with the faces with the same color, as shown in Fig.\,\ref{fig_funcsat_split}. By formulating this local relationship as SAT constraints, the chain of a logical Pauli operator can be moved from initial surface-code patches to the final patches, which constitutes the overall transformation of logical Pauli operators.

Note that while the construction of SAT instances in Ref.~\cite{tan2024} is also based on the idea of stabilizer flows, they treat cubes in the 3D domain as variables. Since our method uses faces and edges of the 3D domain as variables to handle general forms of surface code patches, we re-formulated SAT constraints in a different manner.

For defining Func-SAT, we introduce additional variables, stabilizer face numbers and color variables. 
The binary variables of face numbers are defined as $\mathtt{Num}_{i,j,k,f,b}$, where $f$ is the face index at $(i,j,k)$, and $b$ is the index of the binary number.
In order to check the connectivity between the ports of the index $c$ with the same number and color, the connectivity variables are assigned for each face $i,j,k,f$ as $\mathtt{Connectivity}_{i,j,k,f,c}$. 
The Func-SAT constraint then requires that input and output stabilizer faces with the same color and binary index are connected consistently through local constraints. To enforce path validity, we adopt a SAT encoding of the path-finding problem under the assumption that each node can be connected at most twice.
This SAT constraint enforces consistent stabilizer connectivity:
\begin{equation}
\begin{aligned}
(\lnot \mathrm{Connectivity}_{i,j,k,f,c})\lor \\
[\sum_{p,q=0}^3 (\mathrm{Connectivity}_{S_{p,q},c}\land \mathrm{ConFace}^{i,j,k,f,e_{p,q}}_{S_{p,q}, e'_{p,q}}\wedge\\
\lnot  (\mathtt{FaceColor}_{i,j,k,f} \oplus \mathtt{FaceColor}_{S_{p,q}})\wedge \\
\bigwedge_{b=0}^{N_{\mathtt{qubit}}^{\mathtt{in}}+N_{\mathtt{qubit}}^{\mathtt{out}} }\lnot(\mathtt{Num}_{i,j,k,f,b} \oplus \mathtt{Num}_{S_{p,q},b}))== 2],
\nonumber
\end{aligned}
\end{equation}
where $S_{p,q}$ is the coordinate of the cube and face index, at which the list of the neighbor faces possible connected to the face via the edge $e_{p,q}$ at $(i,j,k,f)$ and the edge $e'_{p,q}$ at $S_{p,q}$.

\subsection{Variables and constraints of FT-SAT}
\begin{figure}[t]
  \centering
  \begin{subfigure}{0.625\columnwidth}
  \includegraphics[width=1\linewidth]{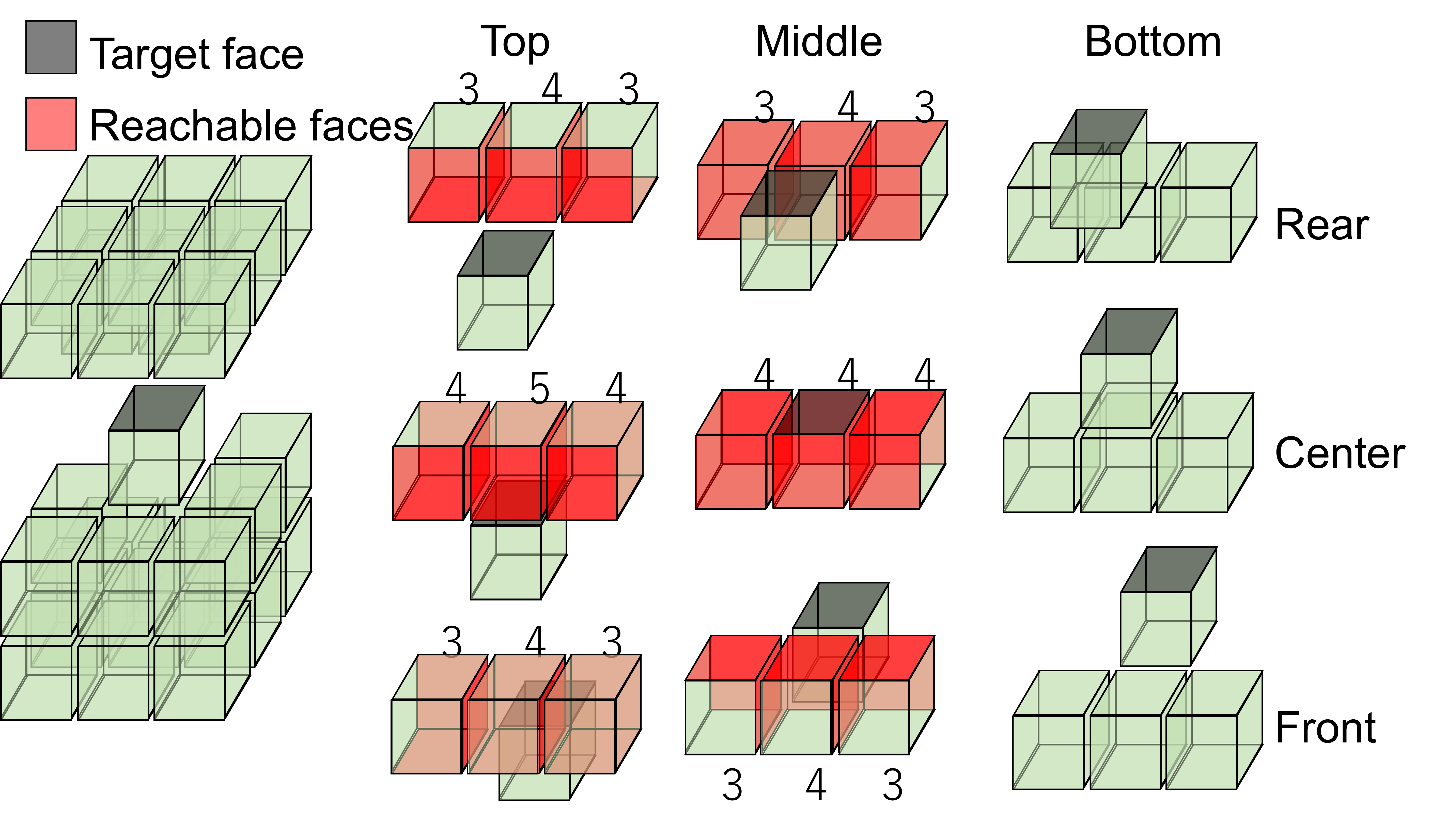}
    
\vspace{-0pt}
    \caption{}
    \label{fig_ftsat_cons}\vspace{-0pt}
  \end{subfigure}\hfill%
  \begin{subfigure}{0.375\columnwidth}
    \includegraphics[width=\linewidth]{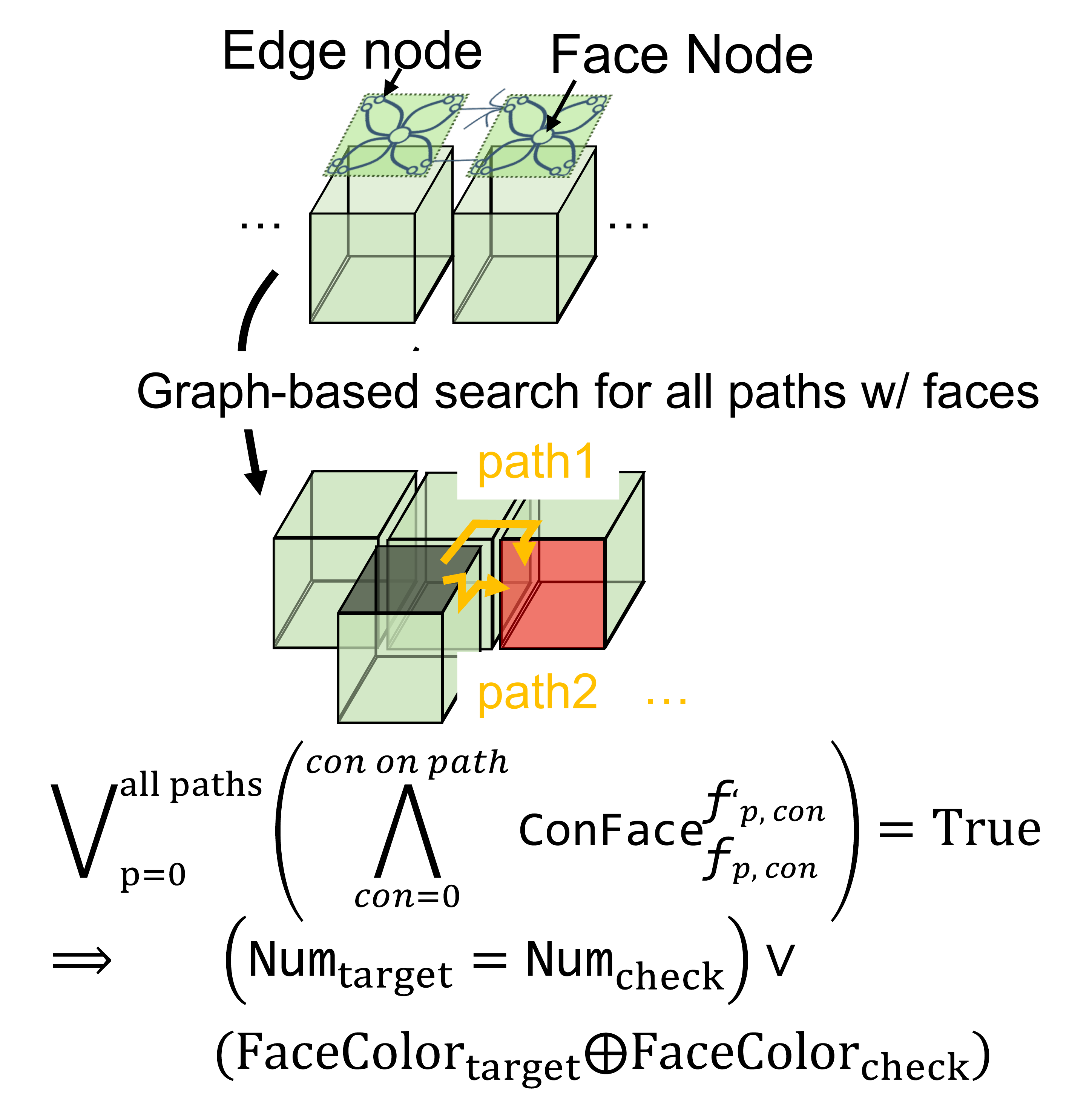}
    
  \vspace{-0pt}
    \caption{}
    \label{fig_ft_exploration}
  \end{subfigure}
  \vspace{-25pt}
  \caption{FT-SAT explanation (a) reachable faces of one target within d-distance (b) exploration of all possible paths corresponding to faces and FT-SAT constraints}
  \label{fig_ft_sat}
  \vspace{-20pt}
\Description{}
\end{figure}

The final constraints, FT-SAT, address the fault tolerance of the bulk.
Since we define faces with a minimum code distance $d$, the fault tolerance of a surface code patch encoding a single logical qubit with a single cell is enforced by the domain discretizations. FT-SAT ensures fault tolerance of code patches encoding multiple logical qubits with multiple patches. 
FT-SAT does not require additional face variables. Instead, it imposes a single constraint: verifying that a pair of connected faces with the same color but with different face numbers does not exist within code distance $d$.
Figure\,\ref{fig_ftsat_cons} illustrates a check example of a pair of faces connected within code distance, i.e., less than one tile length, around a target face. 
The FT-SAT constraint is written as
\begin{equation}
\begin{aligned}
\bigwedge_{p=0}^{N_\mathtt{{reachable\_faces}}}\Bigg( (\lnot \mathtt{FaceExist}_{i,j,k,f} \lor \lnot \mathtt{FaceExist}_{S_p)} \lor \\
(\mathtt{FaceColor}_{i, j, k,f} \oplus \mathtt{FaceColor}_{S_p}) \lor \\ 
(\mathtt{Num}_{i,j,k,f} \oplus \mathtt{Num}_{S_p}) \lor \\
(\bigvee_{q=0}^{N_\mathtt{{Path\_btw.\_face}}}\bigwedge_{con=0}^{N_\mathtt{{ConFace\_on\_path}}}\mathtt{ConFace}_{S_{p,q,con}}^{S'_{p,q,con}}) \Bigg),
\nonumber
\end{aligned}
\end{equation}
where $p$ indexes a face $S_p$ in the checklist for the target face $(i,j,k,f)$, $q$ indexes all paths connecting the two faces within distance $d$, and $con$ indexes each face connection along a path as defined in Sect.~\ref{subsec:LS-SAT}.

To enumerate all possible faces and paths for each target face, we construct a weighted graph over faces and edges in a $3 \times 3 \times 3$ cube domain.
The face distribution and admissible connections in this graph exactly follow those in Sect.~\ref{subsec:LS-SAT}.
Both faces and edges are treated as nodes: a face node connects to its incident edge nodes with weight 0.5, corresponding to half the code distance, and an edge node connects to edge nodes on adjacent faces with weight 0, corresponding to a single ancilla physical used to glue qubit patches.
Starting from the target face, we enumerate all faces reachable via paths of total weight at most 1.
After removing redundant paths using LS-SAT constraints, the total number of such paths to the 65 reachable faces is on the order of $10^3$ per face of a cube.
These constraints are then imposed for all cubes in the exploration domain.

Notice that the above formulation assumes that the code distance $d$ is maintained throughout the entire execution.
In practice, however, several subroutines in FTQC architectures, such as magic-state distillation and entanglement distillation, temporarily use logical qubits with smaller code distances~\cite{litinski2019magic,pattison2025constant} to store noisy quantum information.
Suppose that a certain region needs to maintain code distance $md$, where $m$ is an integer, the above rules are modified to ensure that no pair of connected faces with the same color but with different face numbers appears within a code distance of $md$.
KOVAL-Q can accommodate such situations at the cost of additional constraints, since more faces must be checked, e.g., through connection with total weight up to $m$.

\section{Validation of KOVAL-Q kernel}

\begin{figure}
    \centering
    \includegraphics[width=0.45\linewidth]{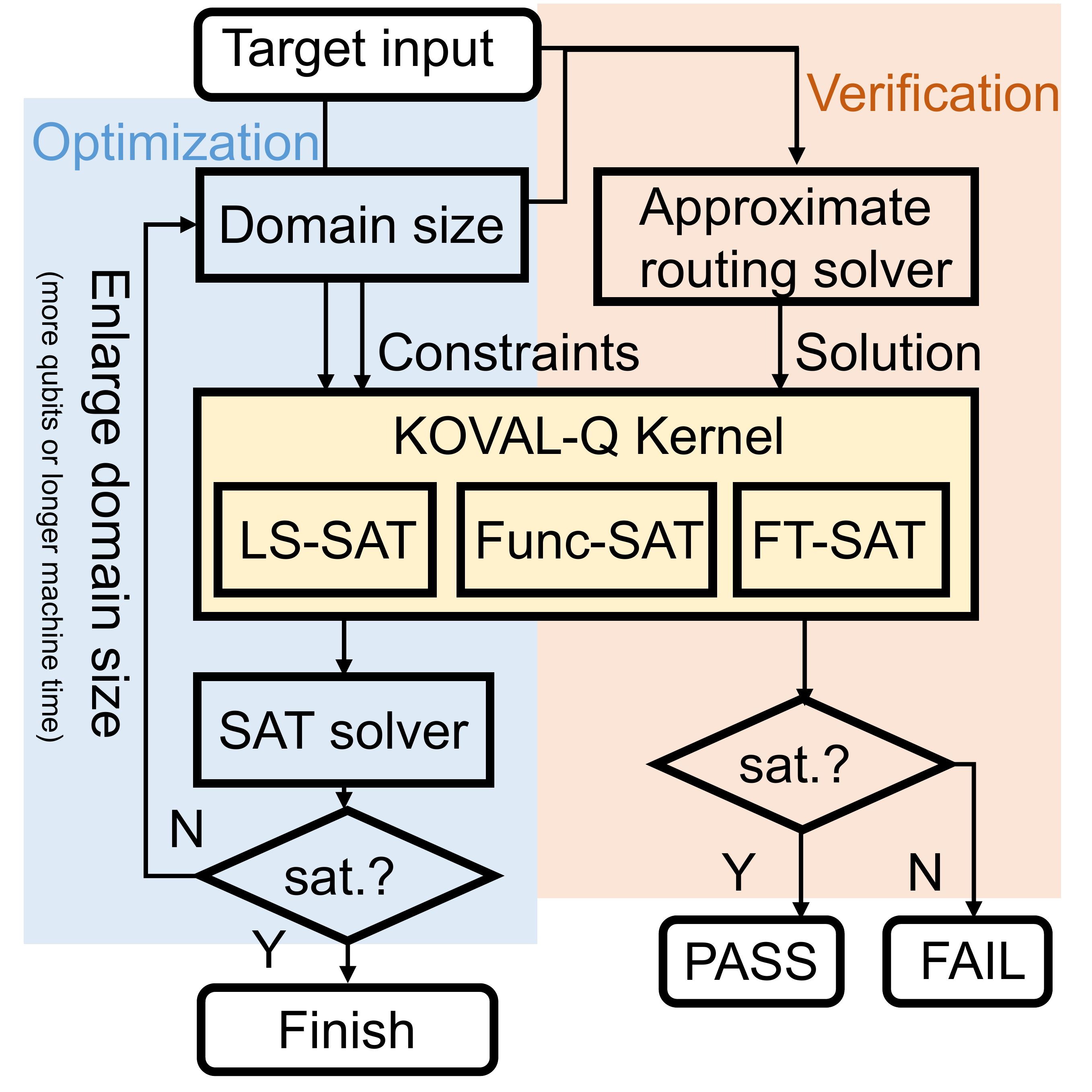}
\vspace{-10pt}
    
    \caption{KOVAL-Q kernel in optimization and verification}
    \label{fig_usage}
    \vspace{-15pt}
\Description{}
\end{figure}

\begin{figure}[!t]
    \centering
    \includegraphics[width=0.6\linewidth]{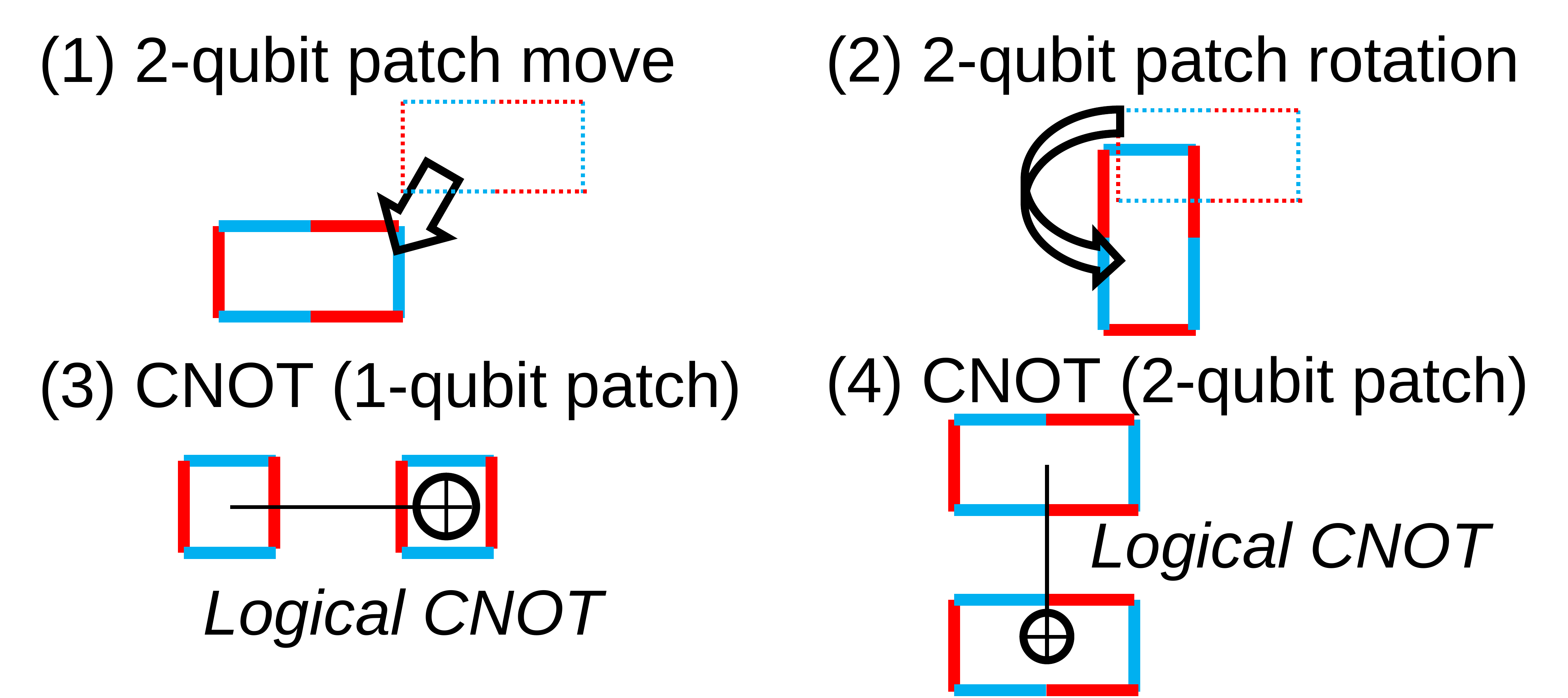}
    
\vspace{-10pt}
\caption{Initializing circuits for verification}
    \label{fig_verification_circuits}\vspace{-15pt}
\Description{}
\end{figure}

KOVAL-Q kernel usage is shown in Fig.\,\ref{fig_usage} covering optimization and verification.
We validate the kernel by finding logical operations based on lattice-surgery for movement, swap, and CNOT gates on 1-qubit and 2-qubit surface-code patches, as shown in Fig.\,\ref{fig_verification_circuits}. 
The test cases are sufficient to demonstrate the feasibility of our method for arbitrary encoded logical qubits. While our current test cases are relatively small domains of size $10^2$ compared to the $10^3$ scale in Ref.\,\cite{tan2024}, this is necessary overhead for handling more general forms of surface-code patches. 
In the final subsection, we will show that this overhead is worthwhile, since KOVAL-Q can even optimize the CNOT operation, one of the most frequently used operations in FTQC, to achieve a twofold reduction in execution time by exploiting multiple-qubit-patch support.

Furthermore, such overhead can be less significant since KOVAL-Q also provides a foundational framework for a universal verification tool for heuristic solvers and FTQC compilers. 
When used in this role, it is analogous to validation stages in classical EDA toolchains, such as design rule checking (DRC) and static timing analysis (STA), in that it validates whether a given candidate implementation satisfies required constraints, avoids global search over the solution space, unlike SAT-based optimization, and thereby substantially alleviates the associated scalability challenges.
We report the number of clauses and variables to illustrate the scalability of the SAT instances themselves. 
Our formulation shows linear growth with domain size and target qubit count, highlighting the potential of KOVAL-Q as a foundation for future EDA tools in FTQC.

\subsection{Convert target to SAT variable values}
\begin{figure}
    \centering
    \includegraphics[width=0.7\linewidth]{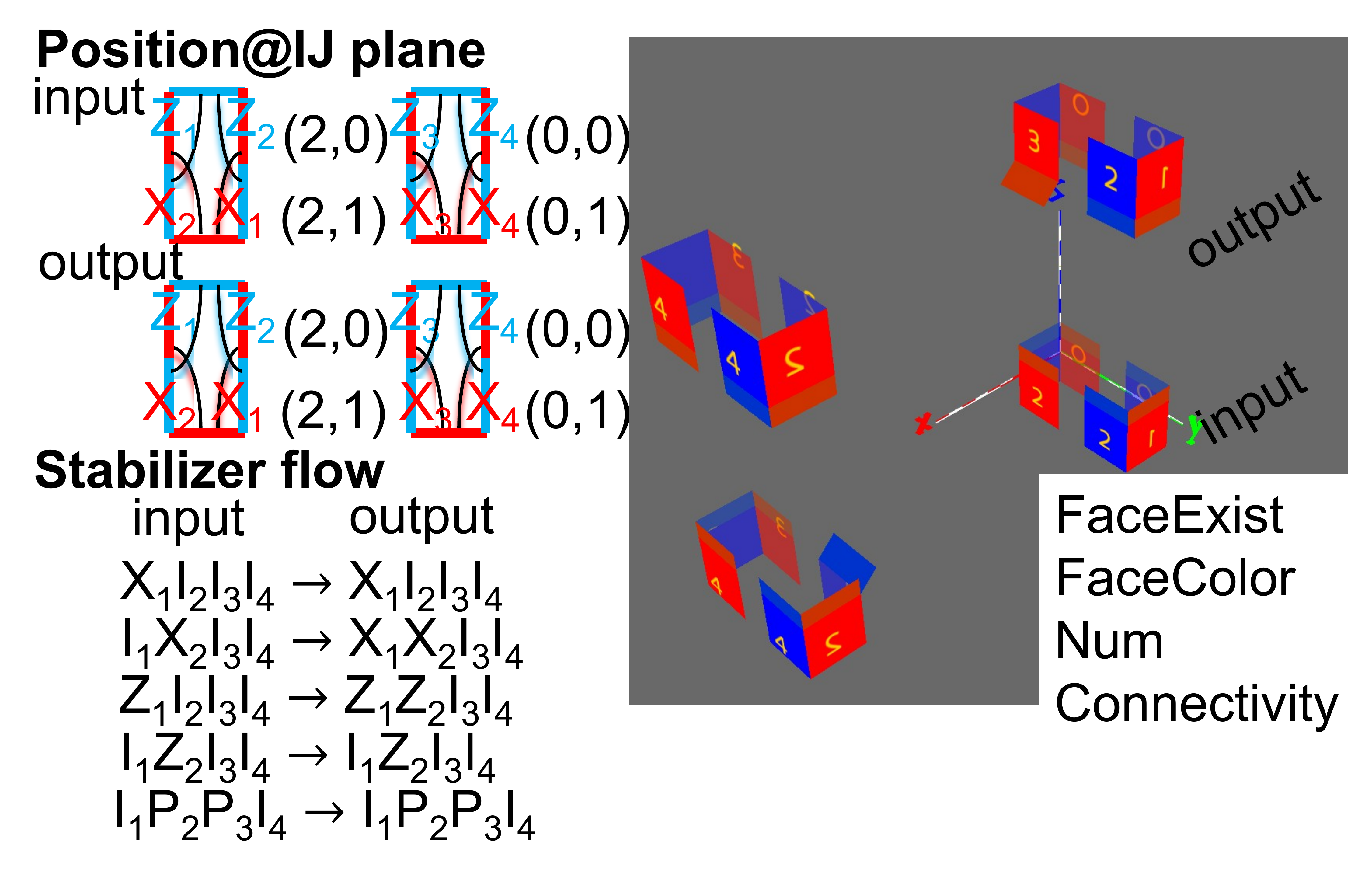}
    
    \vspace{-10pt}
    \caption{Initialization of SAT variables of the CNOT gate}
    \label{fig_initial_example}\vspace{-15pt}
\Description{}
\end{figure}

To generate a SAT instance for a target FTQC implementation with KOVAL-Q, the domain of available resources, target qubit positions, and target logical action must be determined. 
The exploration-domain size of KOVAL-Q is determined by the available device qubits and time resources. 
For a fixed domain size, the set of SAT constraints is always the same except for those related to Func-SAT.
The qubit position information is a target-dependent input, which specifies the positions of initial and final code patches in the domain.
In more detail, they are used to set the existence of stabilizer faces at the top and bottom layers of the domain as input and output ports.
The target logical action provides the stabilizer-flow description of the action, as described in Sec.\,\ref{subsect_funcsat}. 
The connectivity and face-group number variables at the port faces are initialized according to this description. 
Together, these initial values ensure that the generated SAT instance conforms to the constraints of the target FTQC implementation.
Figure\,\ref{fig_initial_example} shows an example for a CNOT gate using two two-qubit patches, along with the corresponding encoding, circuit information, and initialized SAT variables.

\subsection{Synthesizing logical operations acting on multi-qubit patches}
\begin{figure}[t]
  \centering
  \begin{subfigure}{0.5\columnwidth}
  \includegraphics[width=\linewidth]{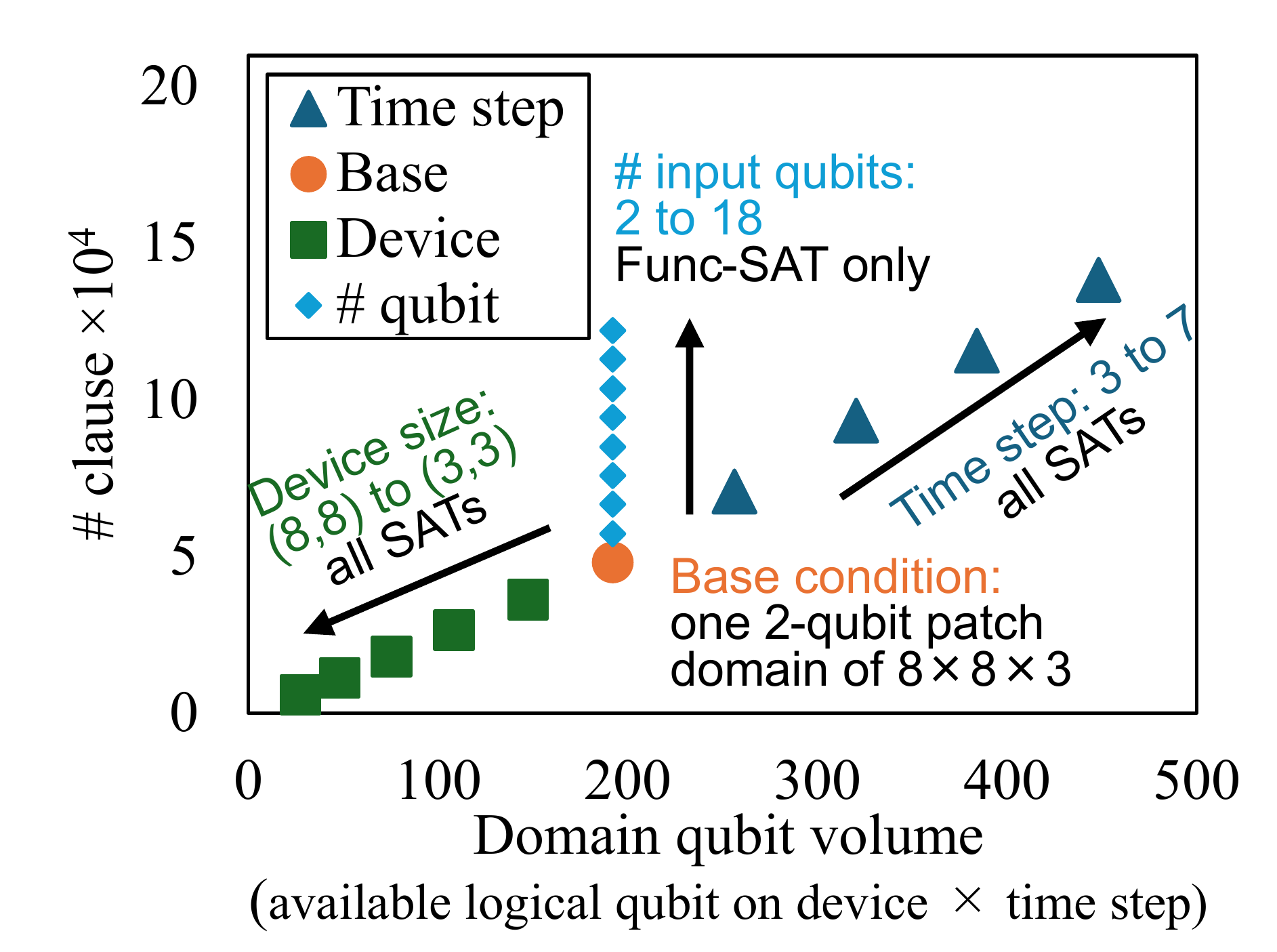}
    
  \vspace{-5pt}\caption{}
    \label{fig_scala_clause}
  \end{subfigure}\hfill%
  \begin{subfigure}{0.5\columnwidth}
    \includegraphics[width=\linewidth]{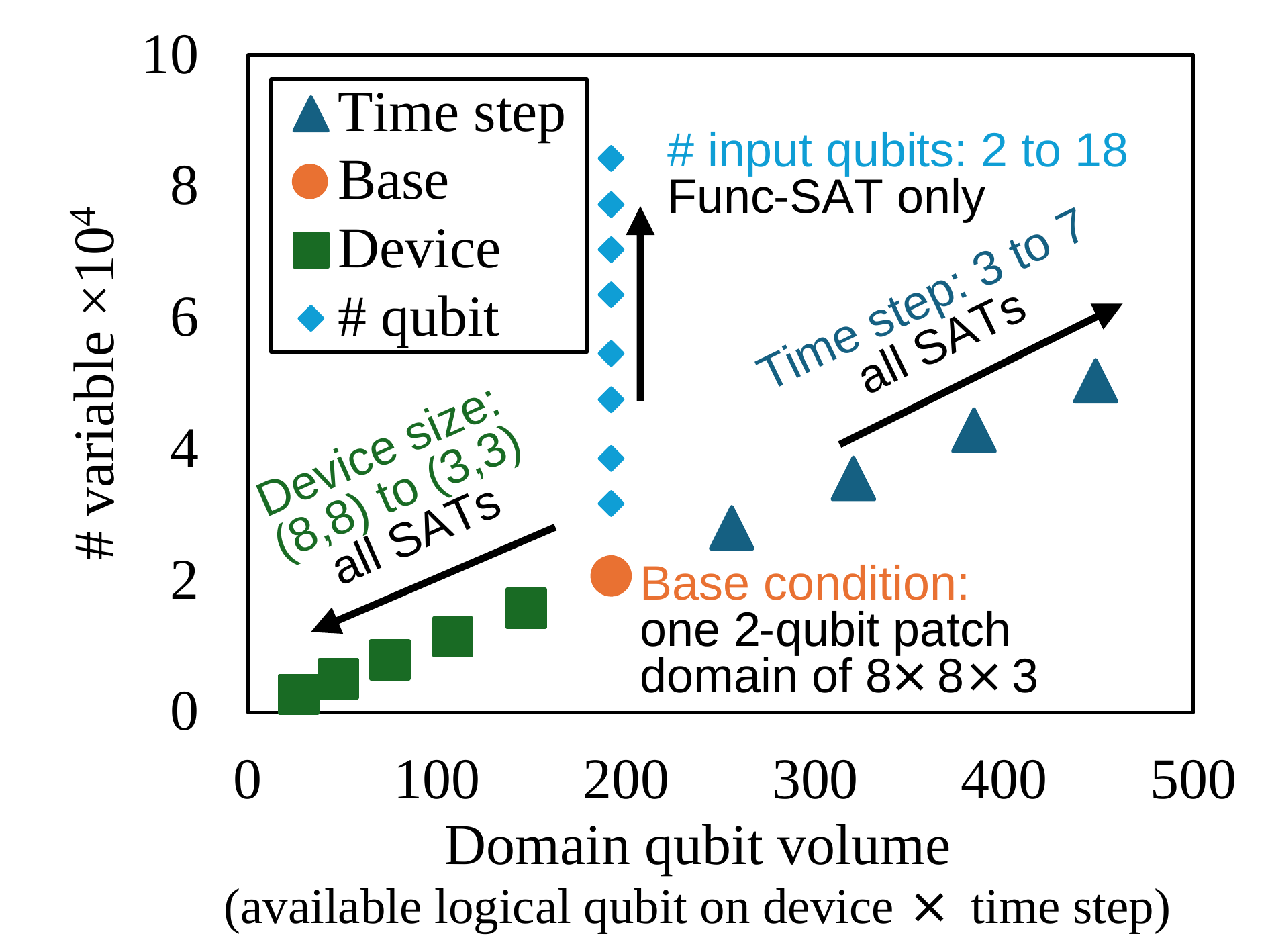}
    
  \vspace{-5pt}
    \caption{}
    \label{fig_scala_vars}
  \end{subfigure}
  \vspace{-20pt}
  \caption{Scalability of (a) clause and (b) variable numbers}
  \label{fig_scalability}
  \vspace{-10pt}
\Description{}
\end{figure}

We tested KOVAL-Q under four test cases listed in Table\,\ref{tab_results}, together with the corresponding domain sizes and SAT instance scales. The device size in the IJ plane was fixed, and we searched for the minimum height along the K axis, which represents machine time in beat units.
The initial time was set to three.
If the SAT instance was unsatisfiable, one extra layer was inserted. 
The first satisfied domain size is used as the minimum time.
The SAT solver was evaluated using Python 3.9 and the z3 solver on a single thread of an Intel Core i9-13900KF with $128~\mathrm{GB}$ of memory. 
Table\,\ref{tab_results} also lists LaSsynth\cite{tan2024} algorithm as a reference.
Although KOVAL-Q has produced larger SAT instances and runs more slowly, it can support arbitrary surface-code encodings.
Figure\,\ref{fig_scalability} further shows the scalability of SAT instance of KOVAL-Q for the movement test case. 
No SAT instance simplification was applied in this experiment.
The results show clear linear scaling with both domain size and qubit count.

\begin{figure}
    \centering
    \includegraphics[width=0.8\linewidth]{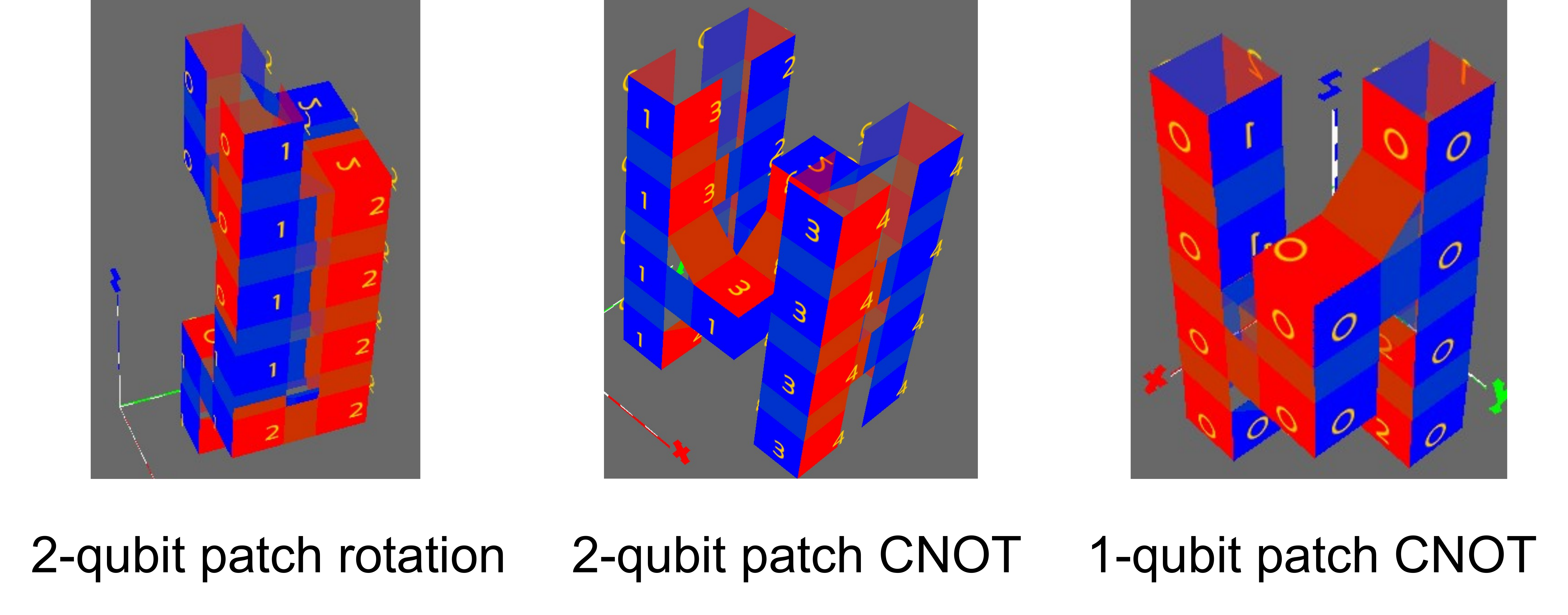}
    
\vspace{-10pt}
    \caption{Examples of visual results of SAT solution}
    
\vspace{-15pt}
\label{fig_SAT_visualization_example}
\Description{}
\end{figure}

Figure\,\ref{fig_SAT_visualization_example} illustrates examples of visualized 3D bulk generated by KOVAL-Q, including rotation and a CNOT operation using two-qubit patches and CNOT operation using one-qubit patches . 
We manually cross-checked that all FTQC implementations automatically generated by KOVAL-Q in the performance and scalability evaluation satisfied the lattice-surgery rules and were functionally correct and fault-tolerant.

\sisetup{group-digits=false}
\begin{table}[t]
  \centering
  \caption{Validation conditions and results}
  \vspace{-10pt}
  \setlength{\tabcolsep}{3pt}
  \renewcommand{\arraystretch}{1}
  \footnotesize
  \begin{tabular}{
    c                     
    S[table-format=1.0]   
    c                     
    c                     
    c                     
    c   
    c   
    c   
  }
    \toprule
    \multirow{1}{*}{KOVAL-Q}& \multicolumn{2}{|c|}{Target input} & \multicolumn{2}{c|}{Domain (IJ $\times$ K)} & \multicolumn{1}{c|}{\#} & \multicolumn{1}{c|}{\#} & {Solver} \\ \cline{2-3} \cline{4-5}
    test case& \multicolumn{1}{|c|}{\#~qubit} & \multicolumn{1}{c|}{Encoding} & \multicolumn{1}{c|}{Dev. } & \multicolumn{1}{c|}{min. time} & \multicolumn{1}{c|}{variables} & \multicolumn{1}{c|}{clauses}& time (s)\\ \hline \midrule
    
    Mov.      & 2 & 2-qubit & $2{\times}2$ & 3 &  1856 &  3725 &  0.24 \\
    Rotation & 2 & 2-qubit & $2{\times}2$ & 8&  3312 &  8052 & 3.5 \\
    CNOT     & 2 & 1-qubit & $2{\times}2$ & 4& 2400 &  4692 &  0.82 \\
    CNOT     & 4 & 2-qubit & $3{\times}2$ & 4& 4516 & 8174 & 3.4 \\
    \toprule
    \multicolumn{5}{l}{Reference results using LaSsynth\cite{tan2024}} &  &  &  \\ 
    CNOT & 2 & 1-qubit & $2{\times}2$ &  4& 280 & 1314 & 0.003 \\
    CNOT & 4 & 1-qubit & $3{\times}2$ &  4& 908 & 6958 & 0.03 \\ 
    CNOT & 4 & 2-qubit & \multicolumn{5}{c}{unsupported}\\
    \bottomrule
  \end{tabular}

\vspace{-10pt}
  \label{tab_results} 
\end{table}

\subsection{Improving logical operations acting on single-qubit patches}

Next, we show that KOVAL-Q can also be used to optimize logical operations acting on logical qubits encoded in single-qubit patches. In particular, we demonstrate that rotations, CNOTs, and qubit movements for logical qubits encoded in a single logical qubit can be improved by expanding the search space with KOVAL-Q.

\begin{figure}
  \centering
  \begin{subfigure}{0.55\columnwidth}
  \includegraphics[width=\linewidth]{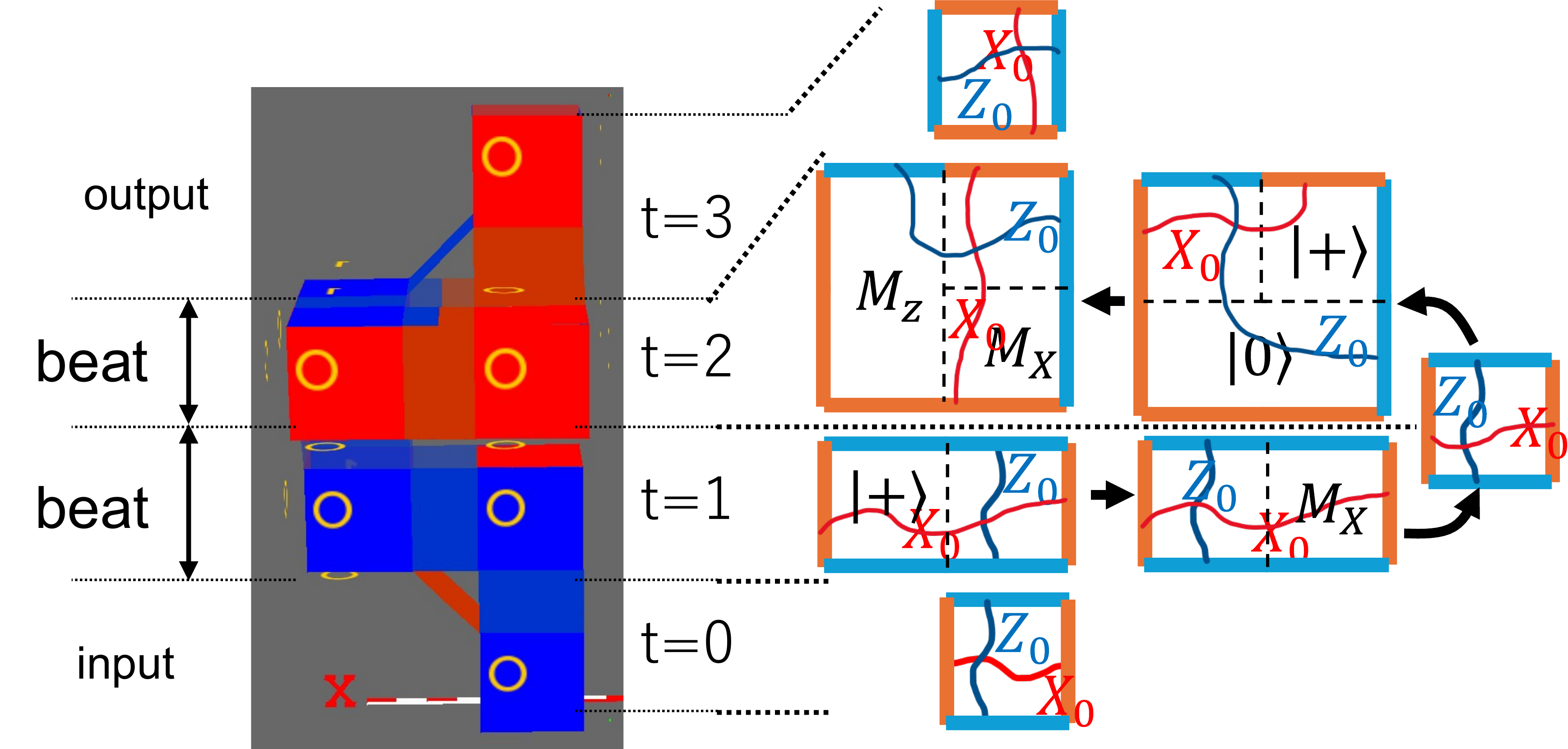}
    
  \vspace{-5pt}\caption{}
    \label{fig_2-beat_rotation}
  \end{subfigure}
\hspace{0.00\columnwidth}
  \begin{subfigure}{0.35\columnwidth}
    \includegraphics[width=\linewidth]{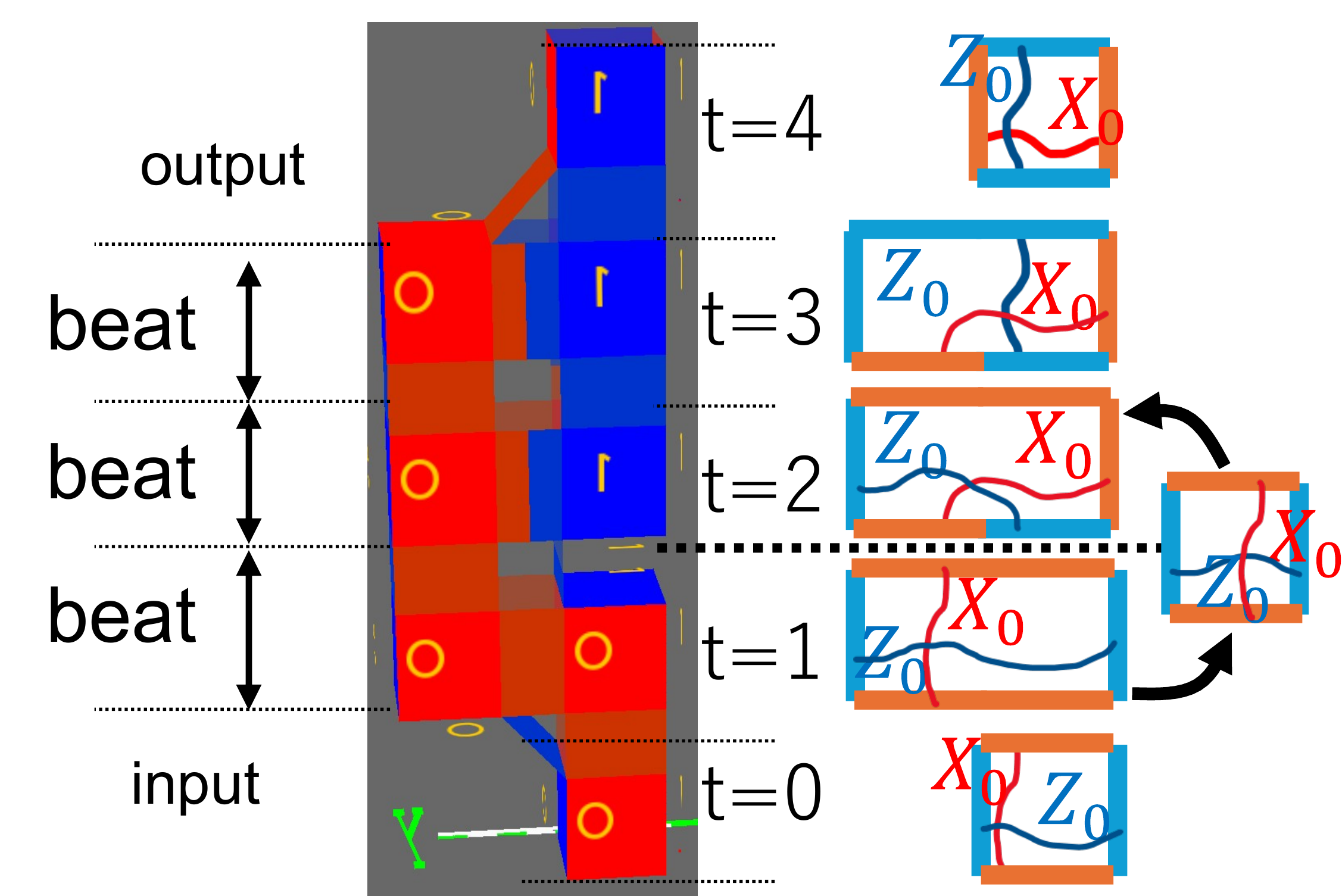}
    
  \vspace{-5pt}
    \caption{}
    \label{fig_3-beat_rotation}
  \end{subfigure}
  \vspace{-10pt}
  \caption{Single-qubit patch rotation: (a) a 2-beat operation in $2\times2$ synthesized by KOVAL-Q; (b) the previously known 3-beat optimal construction in $1\times 2$ \cite{litinski2019game}, also satisfiable in KOVAL-Q. Both of them are unsatisfiable in LaSsynth. Note that 1 beat equals $d$ cycles.}
  \label{fig_patch_rotation}
  \vspace{-20pt}
\Description{}
\end{figure}

A reader might think that single-qubit-patch encodings have already been fully covered by existing studies, such as LaSsynth~\cite{tan2024}, and therefore leave no room for improvement. However, many frameworks, including LaSsynth, implicitly make the following assumptions.
\begin{enumerate}
    \item Assumption 1: 3D diagram is composed of connected $1 \times 1 \times 1$ cubes, where a unit $1$ means $d$ physical qubits in space axis and $d$ cycles for time axis. Each cube is surrounded by two red faces and two blue faces.
    \item Assumption 2:A logical operation is constructed by determining whether each cube is occupied and whether adjacent cubes are connected by pipes.
\end{enumerate}
On the other hand, it is known that efficient and fault-tolerant operations can be generated by going beyond the above assumptions. For example, a logical Hadamard operation in the surface code rotates the boundaries of a qubit by 90 degrees, and therefore the cell of the logical qubit must eventually be rotated by 90 degrees again to restore the original orientation. This operation was first shown in Ref.~\cite{fowler2018low} to be achievable in three beats within a three-patch region. Later, Ref.~\cite{litinski2019game} improved this to two patches and three beats~(Fig.~\ref{fig_3-beat_rotation}). Here, the former construction lies within the search space of LaSsynth, whereas the latter cannot be handled by LaSsynth because its intermediate state contains a boundary configuration that violates assumption 1. In contrast, because KOVAL-Q takes stabilizer faces as its primitive elements, both operations are included in its search space. 

By further expanding the available area to a $2 \times 2$ region, we found that the above execution can be made satisfiable only in two beats. As shown in Fig.~\ref{fig_2-beat_rotation}, by allowing an intermediate $2 \times 2$ configuration with a special boundary, the patch arrangement can be rotated while preserving fault tolerance. To the best of our knowledge, the shortest known rotation operation that does not move the patch position requires three beats, so this result represents a 33\% improvement. Since patch rotation is one of the most fundamental gates in FTQC, this improvement also has a substantial impact on application runtime, as we will evaluate later.

\begin{figure}
  \centering
  \begin{subfigure}{0.4\columnwidth}
  \includegraphics[width=\linewidth]{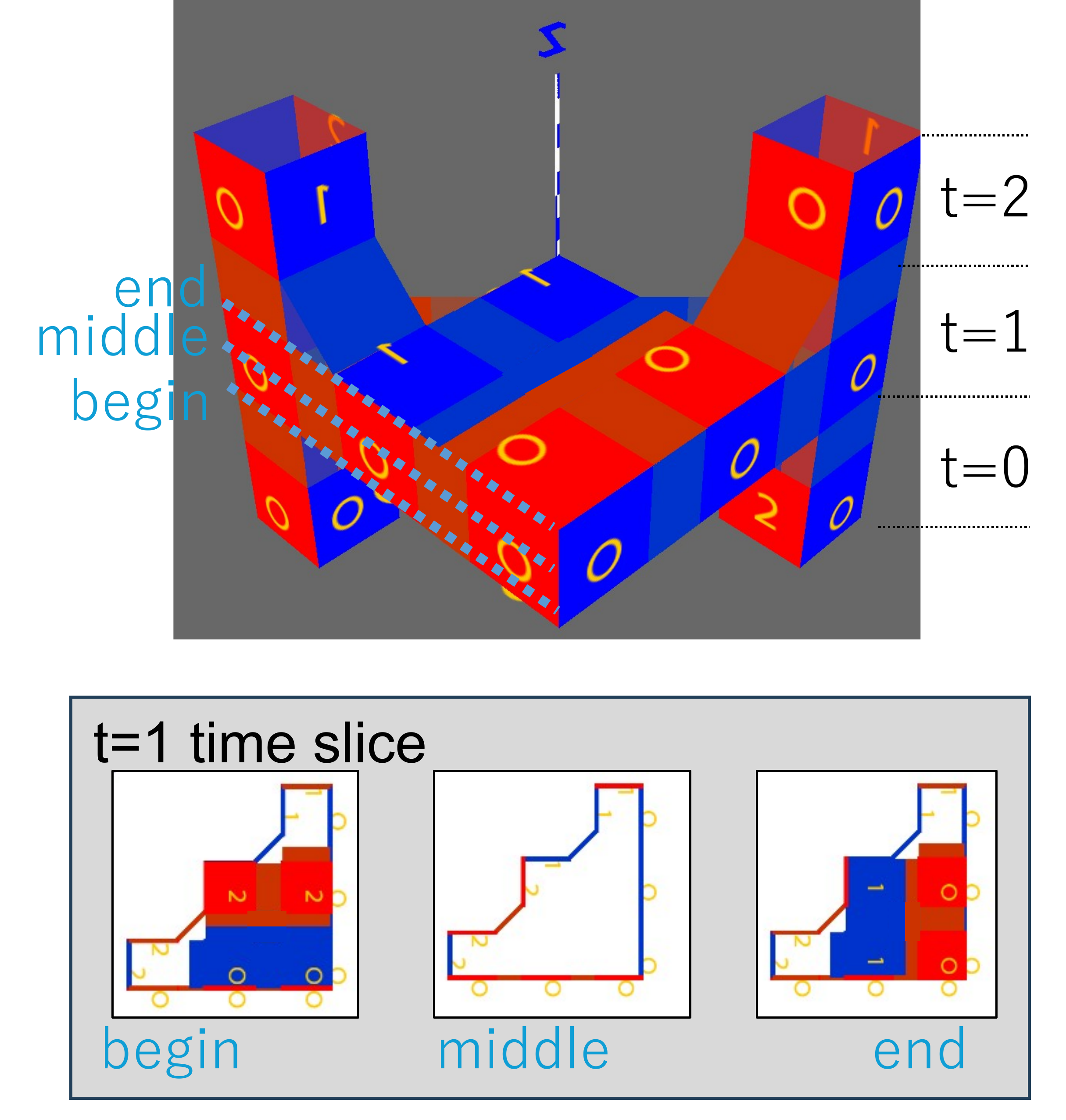}
    
  \vspace{-5pt}\caption{}
    \label{fig_1-beat_cnot_3d}
  \end{subfigure}\hfill%
    \begin{subfigure}{0.5\columnwidth}
    \includegraphics[width=\linewidth]{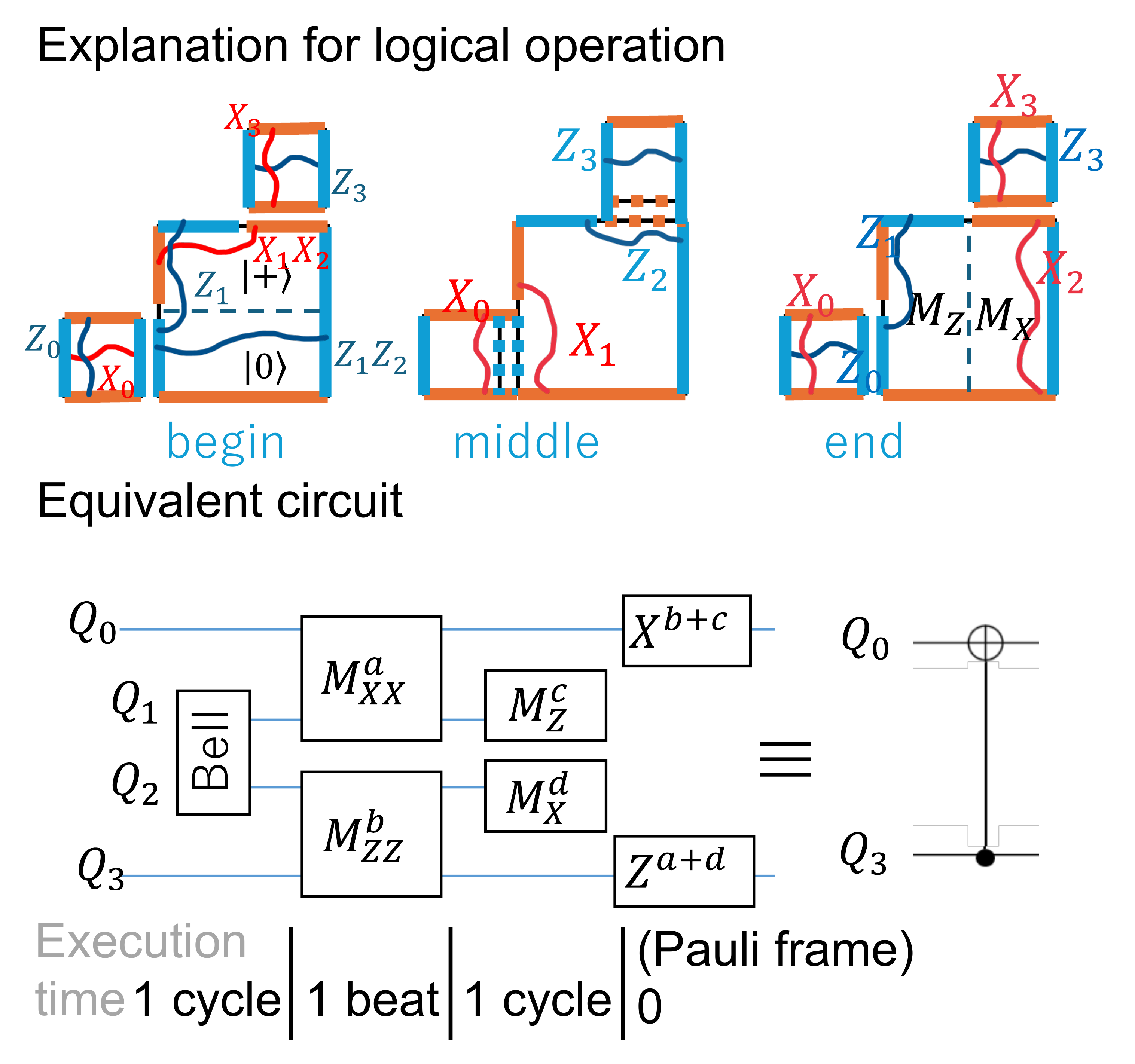}
    
  \vspace{-5pt}
    \caption{}
    \label{fig_1-beat_cnot_slice}
  \end{subfigure}
  \vspace{-10pt}
  \caption{Lattice surgery operation of 1-beat CNOT using 2-qubit patch (a) in 3D block view and 2D time-slice view and (b) its explanation.}\vspace{-17pt}
  \label{fig_1-beat_cnot}

\end{figure}

The same idea can also be applied to the CNOT operation between two logical qubits. In lattice-surgery-based schemes, a CNOT is usually performed in two beats using two lattice-surgery operations, as shown in the right panel of Fig.~\ref{fig_SAT_visualization_example}. In contrast, KOVAL-Q shows that when the two patches are placed with a $2 \times 2$ area between them, for example, at $(0,0)$ and $(2,2)$, the CNOT can be implemented in one beat with the aid of an additional auxiliary patch. The 3D visualization of our construction is shown in Fig.~\ref{fig_1-beat_cnot_3d}. 

The validity of this operation can be understood as follows, as shown in the top panel of Fig.\,\ref{fig_1-beat_cnot_slice}. First, an auxiliary qubit encoded in a $2 \times 2$ multi-qubit patch is created between the two patches. If we choose the logical Pauli basis appropriately, as shown in the figure, the initial state is the $+1$ eigenstate of $Z_2Z_3$ and $X_2X_3$, and this property remains unchanged after the stabilizer measurements. Therefore, we can consider the auxiliary patch to be initialized in a logical Bell state. Furthermore, immediately after this initialization, lattice surgery can be performed fault-tolerantly with the target logical qubits. Finally, the logical qubits are measured in the basis of $Z_1$ and $X_2$, which can be done by measuring all the physical qubits in the patch individually. The resultant quantum circuit is shown in the bottom part of Fig.~12(b). This circuit implements a CNOT with appropriate Pauli-frame feedback, of which the latency can be neglected by using Pauli frame~\cite{riesebos2017}. To the best of our knowledge, no method for performing a CNOT in one beat in this manner has previously been known, and when sufficient space is available, the CNOT latency can therefore be reduced by 50\%.

It is known that a logical CNOT can also be used to move a qubit position to $Q_0$ by initializing $Q_0$ in the $\ket{0}$ state and measuring $Q_3$ in the $X$ basis. Therefore, moving a logical qubit from position $(0,0)$ to $(2,2)$ without rotation, when a $2 \times 2$ auxiliary region is available, would normally require two beats, whereas KOVAL-Q reduces this time to one beat.

Finally, we evaluate the effectiveness of the above improvements in practical settings. In addition to random Clifford circuits, we consider four types of benchmark circuits: ModAdder circuits~\cite{gidney2017factoring}, PREPARE and SELECT circuits~\cite{babbush2018}, and Trotter-simulation circuits. These benchmarks represent subroutines arising in integer factorization, ground-state energy estimation, and time-evolution simulation, respectively. We generated these instances using Quration~\cite{suzuki2026quration} and converted them into the Clifford+T formalism using gridsynth~\cite{ross2014optimal}. The circuits were synthesized such that the number of logical qubits ranges from 16 to 32 and the number of instructions exceeds 2000. A baseline scheme uses 3-beat patch rotations after Hadamard gates and CNOT gates are decomposed into consecutive two lattice-surgery operations. The proposed scheme is allowed to use 2-beat patch rotations and 1-beat CNOT gates. For simplicity, we counted execution time along the critical paths of the instruction dependency graph, i.e., ignoring routing conflicts. We also assumed that a sufficient number of magic states are supplied. Although $S$ gates are randomly inserted after $T$-gate teleportation in general, we assumed that they are always inserted for simplicity.

Based on these assumptions, we measured the execution time in units of code beats and compared the results with the baseline. For random Clifford circuits, we observed a 22\% improvement over the baseline. For the application instances, we observed improvements of 9\%, 6\%, 11\%, and 14\% for the ModAdder, SELECT, PREPARE, and Trotter-simulation circuits, respectively. Therefore, the subroutines generated by KOVAL-Q provide substantial improvements even in cases where concrete implementations were already known.

\section{RELATED WORK}
\label{sec:bgd}
Establishing a universal method for finding fault-tolerant implementations of quantum circuits based on given QEC codes is a major goal in the development of FTQC. In this work, we refer to the exploration of optimized FTQC implementations as EDA for FTQC. To the best of our knowledge, no prior work has established a general foundation for an FTQC EDA kernel compatible with general surface-code encodings. In this work, we exploit the fact that the properties of surface-code encodings and lattice-surgery operations can be reformulated in topological terms. Based on this observation, we propose a general kernel formulated as geometric constraints in a Boolean SAT problem.

Prior work on FTQC EDA tools is summarized in Table~\ref{table_comparison}. These works can be grouped into several classes. The first class~\cite{litinski2019game,litinski2019magic,beverland2022assessing,watkins2024high,kobori2025lsqca,hamada2024efficient} fixes the placement and encoding method of logical qubits on the device plane. This strategy simplifies the rules of lattice surgery, since FTQC implementations can then be reduced to problems that admit efficient solvers, such as path-finding problems. The main advantage of this approach is its high efficiency, as many classical routing algorithms can be used to schedule large numbers of lattice-surgery operations in utility-scale programs. However, although several recent works~\cite{kan2025,trochatos2025} have begun to support multiple architectures, their dependence on a limited set of specific architectures still restricts their portability across different FTQC platforms. Moreover, because qubit placement and implementation segmentation are fixed, global optimization is difficult to achieve.

The second class consists of hand optimization. Some important logical subroutines, such as CNOT gates and magic-state generation~\cite{fowler2018low,litinski2019magic,litinski2019game}, have been manually optimized using layouts tailored to particular device architectures. Although such designs are known to be efficient, it is not practical to optimize every subroutine for all given encoding schemes and hardware configurations. It is also difficult to verify whether a derived solution is optimal within the considered exploration space.

Recently, LaSsynth~\cite{tan2024} was proposed as an SAT-based approach for globally optimizing FTQC subroutines. In this framework, however, the encoding of each logical qubit is strictly limited to a single-qubit surface-code patch. Although this simplification reduces SAT complexity and enables applications such as magic-state factories, it is incompatible with general surface-code encodings that are widely used in state-of-the-art architectures~\cite{litinski2019game,beverland2022assessing,silva2024multi,toshio2026star}.

In contrast, our proposed KOVAL-Q emphasizes universality over simplicity of constraints. Although SAT is NP-hard in general, KOVAL-Q can optimize various types of small but important subroutines for general forms of surface codes, and it is also sufficiently fast when used to validate given solutions. We therefore focus on developing a universal EDA kernel that is applicable across architectures based on surface codes and potentially extendable to other topological stabilizer codes, such as color codes~\cite{landahl2011fault,lacroix2025scaling}. KOVAL-Q is designed to serve both as a reusable kernel for optimizing FTQC subroutines and as a verification tool for full implementations. For example, although the scalability of LaSsynth is also limited, Ref.~\cite{hao2025compilation} uses LaSsynth as a kernel to optimize subcircuits of whole quantum programs. We expect that methods in the first class can likewise incorporate solutions generated by KOVAL-Q as subroutines.

\begin{table}[t]
  \centering
  \caption{Summary of previous FTQC EDA tools}
\vspace{-10pt}
  \small
\setlength{\tabcolsep}{3pt}
\renewcommand{\arraystretch}{1.}
\resizebox{\columnwidth}{!}{%
  \begin{tabular}{lccc}
    \toprule
    \multicolumn{1}{c|}{\textbf{Previous work}} & \multicolumn{1}{c|}{\textbf{Implement. method}} & \multicolumn{1}{c|}{\textbf{Qubit position}} & \multicolumn{1}{c}{\textbf{Qubit encoding}}  \\  \hline

    \midrule
    Flower\cite{fowler2018low}        & Manual \& Graph-based            & fixed  & 1-qubit \\
    Litinski\cite{litinski2019game}         & Manual \& Graph-based           & fixed  & 1-\& 2-qubit    \\
    Litinski\cite{litinski2019magic}    & Manual     & fixed  & 1-qubit         \\
    Beverland\cite{beverland2022assessing} & Manual \& Graph-based & fixed & 1-\&2-qubit \\
    Silva\cite{silva2024multi} & Graph-based & fixed & 1-\&2-qubit \\
    LaSsynth\cite{tan2024}              & SAT                 & arbitrary & 1-qubit\\
    Molavi\cite{molavi2025}           & Annealing  & fixed, remap poss.  & 1-qubit         \\
    Kan\cite{kan2025}              & Graph-based        & fixed  & 1-qubit         \\
    \midrule
    KOVAL-Q & SAT & arbitrary & arbitrary\\
    \bottomrule
  \end{tabular}}
  \label{table_comparison}
  
  \vspace{-10pt}
\end{table}

\section{Conclusion}
We proposed KOVAL-Q, an EDA kernel for optimizing and verifying logical operations on general surface-code encoding schemes. KOVAL-Q expresses all constraints in terms of stabilizer faces as the fundamental decision variables and formulates the design of fault-tolerant implementations as SAT instances. By framing FTQC synthesis as an EDA kernel rather than as a monolithic optimizer, KOVAL-Q provides a reusable foundation for both subroutine optimization and whole-circuit verification.

Our formalism can handle general surface-code encodings such as multi-qubit patches as target qubits and intermediate states, thereby expanding both the applicability of SAT-based synthesis and the exploration space for optimization and verification. We demonstrated that KOVAL-Q can successfully synthesize fault-tolerant logical Clifford operations on four logical qubits encoded in multi-qubit patches. Moreover, thanks to the expanded exploration space, KOVAL-Q reduces the duration of a logical CNOT gate, qubit movement, and patch rotations for logical qubits encoded in single-qubit patches. Because these are fundamental logical operations, the formulation underlying KOVAL-Q is highly beneficial for achieving near-optimal designs of such operations. Although our formulation introduces overhead in the number of variables and constraints compared with Tan~\textit{et al.}~\cite{tan2024}, this issue can be mitigated by using KOVAL-Q to optimize subcircuits of whole quantum programs, such as logical CNOT gates, in the same way that LaSsynth is used in Ref.~\cite{hao2025compilation}. Therefore, we believe that KOVAL-Q provides an important framework for the efficient design of FTQCs and establishes a foundation for future EDA tools for FTQC.

\ifarxiv
  \section*{Acknowledgments}
  This work is supported by the Grant-in-Aid for Early-Career Scientists from Japan Society for the Promotion of Science (JSPS) under Grant JP24K20755, and the Grant-in-Aid for Scientific Research (S) from JSPS under Grant JP24H00073.
  This work is supported by Ministry of Education, Culture, Sports, Science and Technology (MEXT) Q-LEAP Grant No.~JPMXS0120319794 and JPMXS0118068682, MEXT Feasibility Study on the future HPCI, Japan Science and Technology (JST) Moonshot R\&D Grant No.~JPMJMS2061, JST CREST Grant No.~JPMJCR23I4, JPMJCR24I4, and JPMJCR25I4.
\else
\ConfAcknowledgment
\fi

\bibliographystyle{ACM-Reference-Format}
\bibliography{refs}

\end{document}